\UseRawInputEncoding

\documentclass[]{raa}           

\usepackage{graphicx,times}
\usepackage{natbib}
\usepackage{amssymb,amsmath}
\usepackage{multirow}
\bibpunct{(}{)}{;}{a}{}{,}

\usepackage[pagebackref=true]{hyperref}

\begin{document}

\title{
  Photometric redshift estimates using Bayesian neural networks in the CSST survey
}

\volnopage{Vol.0 (20xx) No.0, 000--000}      
\setcounter{page}{1}          

\author{Xingchen Zhou 
  \inst{1,2}
  \and Yan Gong$^*$
  \inst{1}
  \and Xian-Min Meng
  \inst{1}
  \and Xuelei Chen
  \inst{4,2,5}
  \and Zhu Chen
  \inst{6}
  \and Wei Du
  \inst{6}
  \and Liping Fu
  \inst{6}
  \and Zhijian Luo
  \inst{6}
}

\institute{National Astronomical Observatories, Chinese Academy of Sciences,
Beijing 100101, China; {\it gongyan@bao.ac.cn}\\
\and
University of Chinese Academy of Sciences, Beijing 100049, China;\\
\and
Science Center for China Space Station Telescope, National Astronomical Observatories,
Chinese Academy of Sciences, 20A Datun Road, Beijing 100101, China;\\
\and
Key Laboratory of Computational Astrophysics, National Astronomical Observatories,
Chinese Academy of Sciences, 20A Datun Road, Beijing 100101, China;\\
\and
Center for High Energy Physics, Peking University, Beijing 100871, China;\\
\and
Shanghai Key Lab for Astrophysics, Shanghai Normal University, Shanghai 200234, China
\vs\no
{\small Received 20xx month day; accepted 20xx month day}}

\abstract{
  Galaxy photometric redshift (photo-$z$) is crucial in cosmological studies, such as weak
  gravitational lensing and galaxy angular clustering measurements. In this work, we try to extract
  photo-$z$ information and construct its probability distribution function (PDF) using the Bayesian neural networks (BNN) from
  both galaxy flux and image data expected to be obtained by the China Space Station Telescope (CSST).
  The mock galaxy images are generated from the Advanced Camera for
  Surveys of Hubble Space Telescope ($HST$-ACS) and COSMOS catalog, in which the CSST instrumental
  effects are carefully considered. And the galaxy flux data are measured from galaxy images using aperture
  photometry. We construct Bayesian multilayer perceptron (B-MLP) and Bayesian convolutional neural network (B-CNN)
  to predict photo-$z$ along with the PDFs from fluxes and images, respectively. We combine the B-MLP and B-CNN together, and
  construct a hybrid network and employ the transfer learning techniques to investigate the improvement of
  including both flux and image data.
  For galaxy samples with SNR$>$10 in $g$ or $i$ band, we find the accuracy and outlier fraction of photo-$z$
  can achieve $\sigma_{\rm NMAD}=0.022$ and $\eta=2.35\%$ for the B-MLP using flux data only, and $\sigma_{\rm NMAD}=0.022$ and $\eta=1.32\%$
  for the B-CNN using image data only. The Bayesian hybrid network can achieve $\sigma_{\rm NMAD}=0.021$ and $\eta=1.23\%$, and utilizing
  transfer learning technique can improve results to $\sigma_{\rm NMAD}=0.019$ and $\eta=1.17\%$, which can provide the most confident predictions
  with the lowest average uncertainty.
  }

\authorrunning{X. Zhou et al.}            
\titlerunning{Photo-$z$ by BNN}  

\maketitle

%
%
\section{Introduction}           
\label{sect:intro}
According to current cosmological observations, about 95\% components of our Universe are dark matter and dark energy, far more abundant than
luminous objects. Dark matter and dark energy are major concerns in current cosmological studies, and they leave their footprints
at both small and large scales, such as galaxies and large-scale structure. A number of ongoing and
next-generation surveys attempt to detect these footprints in wide and deep survey areas, e.g. the Sloan Digital Sky Survey~\citep{Fukugita96,York2000},
Dark Energy Survey~\citep{Abbott2016,Abbott2021},
the Legacy Survey of Space
and Time (LSST) or Vera C. Rubin Observatory~\citep{Abell2009,Ivezic2019}, the Euclid Space Telescope~\citep{Laureijs2011}
and the Wide-Field Infrared Survey Telescope or Nancy Grace Roman
Space Telescope~\citep{Green2012,Akeson2019}. These surveys are expected to obtain
huge amount of galaxies with photometric information, such as magnitude, color, morphology, etc. Then powerful
cosmological probes such as weak gravitational lensing and galaxy angular clustering can be accomplished and provide excellent constraint on dark matter,
dark energy and other important objects in the Universe.

Weak lensing (WL) and many other cosmological probes need reliable distance or redshift measurements of large number of galaxies. Accurate galaxy redshifts
can be measured by fitting emission or absorption lines in galaxy spectra. However, obtaining accurate
spectra and redshifts is time-consuming, which is not suitable for current weak lensing observations. \citet{Baum1962} proposed that redshift can be obtained from
far more less time-consuming photometric information, resulting a photometric redshift (photo-$z$). The accuracy of photo-$z$
is one of the main systematics in many cosmological studies including WL. Photo-$z$ is becoming an essential quantity nowadays
and approaches to improve photo-$z$ accuracy is under active research. Two main methods are utilized to derive photo-$z$ given
photometric information. One is template fitting method, where spectral energy distributions (SEDs) are used to fit
photometric data in multi-band to obtain photo-$z$~\citep{Lanzetta96,Fernandez99,Bolzonella2000}.
The other one is deriving empirical relations between photometric data
and redshift from existing data. This method can be called as training method, and mostly is accomplished by Machine
Learning (ML), especially neural networks~\citep{Collister2004,Sadeh2016,Brescia2021}. Both methods have their own advantages.
The template fitting method can efficiently derive photo-$z$ if SED templates are representative enough for the considered samples.
And the training method can obtain more
accurate photo-$z$ if training data have reliable spectroscopic redshifts and are sufficiently large to cover all features
of galaxies in a photometric survey.

Acquiring large amount of galaxies with reliable spectroscopic redshift for training sample is challenging, however, there
are a number of spectroscopic galaxy surveys are ongoing and arranged currently, e.g. Dark Energy Spectroscopic Instrument
~\citep{Levi2019}, Prime Focus Spectrograph (PFS, ~\citealt{Tamura16}), Multi-Object Optical and Near-infrared Spectrograph~\citep{Cirasuolo20,Maiolino20},
4-meter Multi-Object
Spectroscopic Telescope~\citep{deJong2019}, MegaMapper~\citep{Schlegel19}, Fiber-Optic Broadband Optical Spectrograph~\citep{Bundy19}
and SpecTel~\citep{Ellis19}. These
surveys will provide a huge amount of galaxy spectra with spectroscopic redshifts, which can be constructed to be training
samples for machine learning.
Currently, neural network algorithm is mostly under remarkable development among various machine learning
algorithms. Two widely used neural networks are utilized in astronomical and cosmological studies, i.e, multilayer
perceptron (MLP) and convolutional neural network (CNN). The MLP is usually constructed by input layers, several hidden layers and
output layers. Every MLP layer is formed by several computing neurons where weights and biases control the output~\citep{Haykin1994}. The CNN,
introduced by~\citet{Fukushima1982} and ~\citet{Lecun1998}, can extract useful features by multiple learnable kernel arrays and
show great success in computer vision.

At present, most neural networks only give point estimate, but the confidence or uncertainty of prediction is also important in many tasks.
The uncertainties on one hand come from corruption of datasets, such as blurring and measurement errors, on the other hand,
come from networks, because of a large number of learnable weights.
Bayesian neural network (BNN, ~\citealt{Bishop1997,Blundell2015,Gal2015}) can
capture both uncertainties by outputting variance of predictions, and consider weights as posterior distributions learned from
data and priors by Bayesian algorithm. For regression tasks, probability distribution function (PDF) of prediction can also be
constructed by this network. Therefore, Bayesian neural network should be a promising tool in astronomical and cosmological studies,
that can estimate uncertainties or PDFs of important quantities.

\begin{figure}
  \centering
  \includegraphics[width=\textwidth, angle=0]{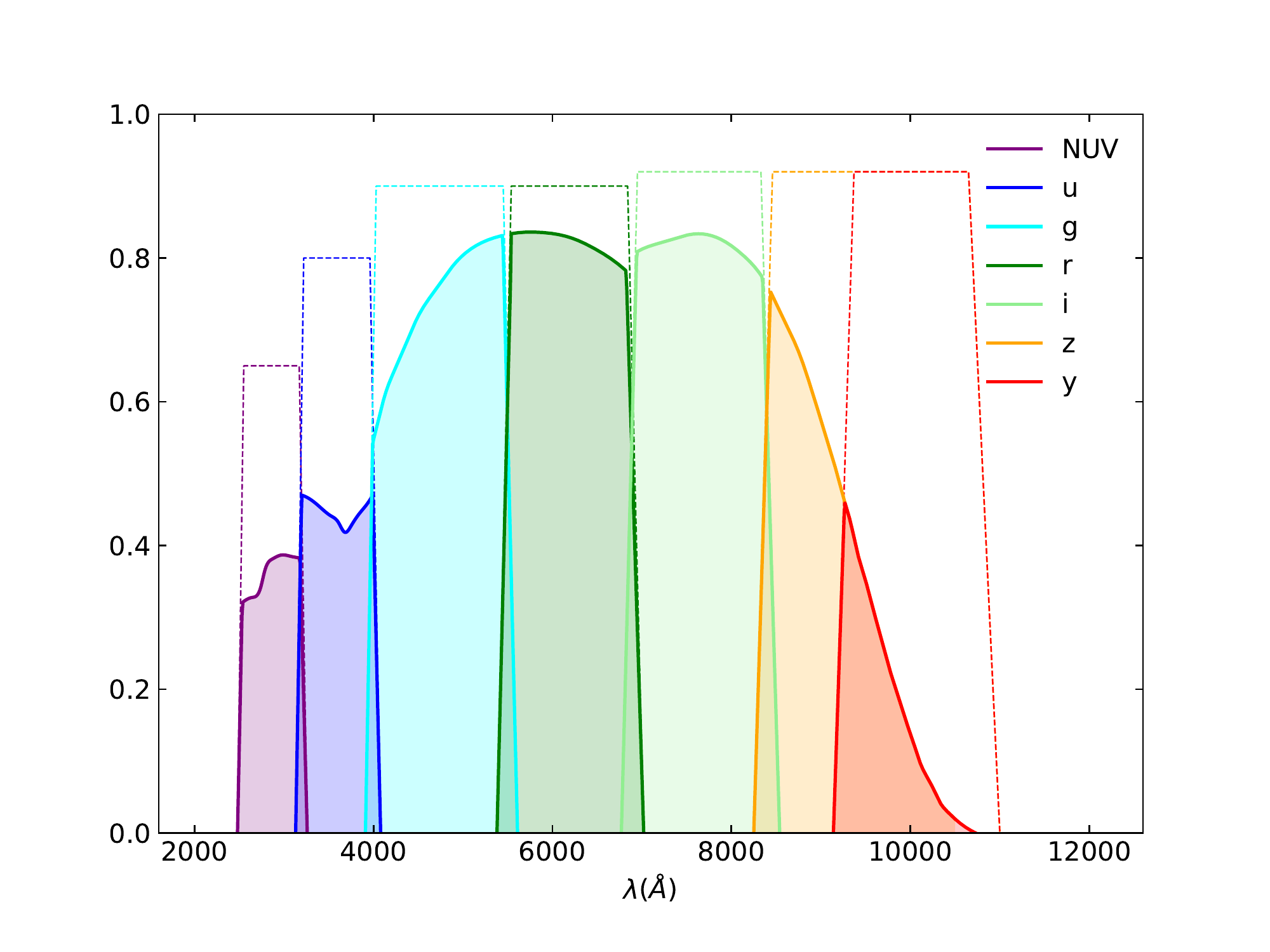}
  \caption{The CSST intrinsic transmissions (dashed) and total transmissions with detector quantum efficiency (solid) of 7 photometric filters.
    The details of parameters of transmissions can be found in~\citep{Cao2018} and Meng et al. (in preparation).}
  \label{fig:transmission}
\end{figure}

In this work, we employ Bayesian neural networks to study the accuracy of photo-$z$ along with its uncertainty or PDF
for China Space Station Telescope (CSST). The CSST, a 2-m space telescope, is scheduled to launch around 2024 and reaches the
same orbit with the China Manned Space Station~\citep{Zhan2011,Zhan2018,Zhan2021,Cao2018,Gong2019}.
It has 7 photometric filters, i.e. $NUV, u, g, r, i, z$ and $y$. These filters
cover wavelength range from $\sim2500$ to $\sim10000$ \AA\ and have $5\sigma$ magnitude limit for point-source detection as $25.4, 25.4, 26.3, 26.0, 25.9,
  25.2$ and $24.4$ AB mag, respectively. Figure~\ref{fig:transmission} shows the intrinsic transmissions and total transmissions considering detector quantum efficiency of 7 filters, and the
details of transmission can be found in~\citep{Cao2018} and Meng et al. (in preparation).
One of the CSST's main probe is weak gravitational lensing survey, which is highly dependent on the accuracy of photo-$z$ measurements. Some of previous studies already research
photo-$z$ using neural networks for the CSST. For example, ~\citet{Zhou2021} uses simple MLP to predict photo-$z$ from mock data directly derived from galaxy
SEDs, and ~\citet{Zhou2022} applies MLP and CNN to predict redshifts from mock flux data and galaxy images. However, these two works only capture uncertainties
partially or just give point estimates for photo-$z$, and in this work we will apply BNN to estimate photo-$z$s and their PDFs as well.

This paper is organized as follows: Section~\ref{sect:mock data} explains generation of the mock images and flux data. In Section~\ref{sect:methods}, we
introduce concepts of Bayesian neural network and present its implementation and details of training process. Results are
illustrated in Section~\ref{sect:results}. Finally we conclude our results in Section~\ref{sect:conclusion}

\section{Mock Data}
\label{sect:mock data}

Mock images are generated based on the F814W band of Advanced Camera for Surveys of \textit{Hubble Space Telescope} ($HST$-ACS)
and COSMOS catalogs~\citep{Koekemoer2007,Massey2010,Bohlin2016,Laigle2016}.
This survey has similar spatial resolution as the CSST, and background noise is $\sim1/3$ of the CSST photometric survey. Therefore,
it provides a good basis to simulate galaxy images of CSST photometric survey as real as possible.
The details of image generation procedure are explained in~\citet{Zhou2022} and Meng et al. (in preparation), and we summarize the important points here.

\begin{figure}
  \centering
  \includegraphics[width=\textwidth, angle=0]{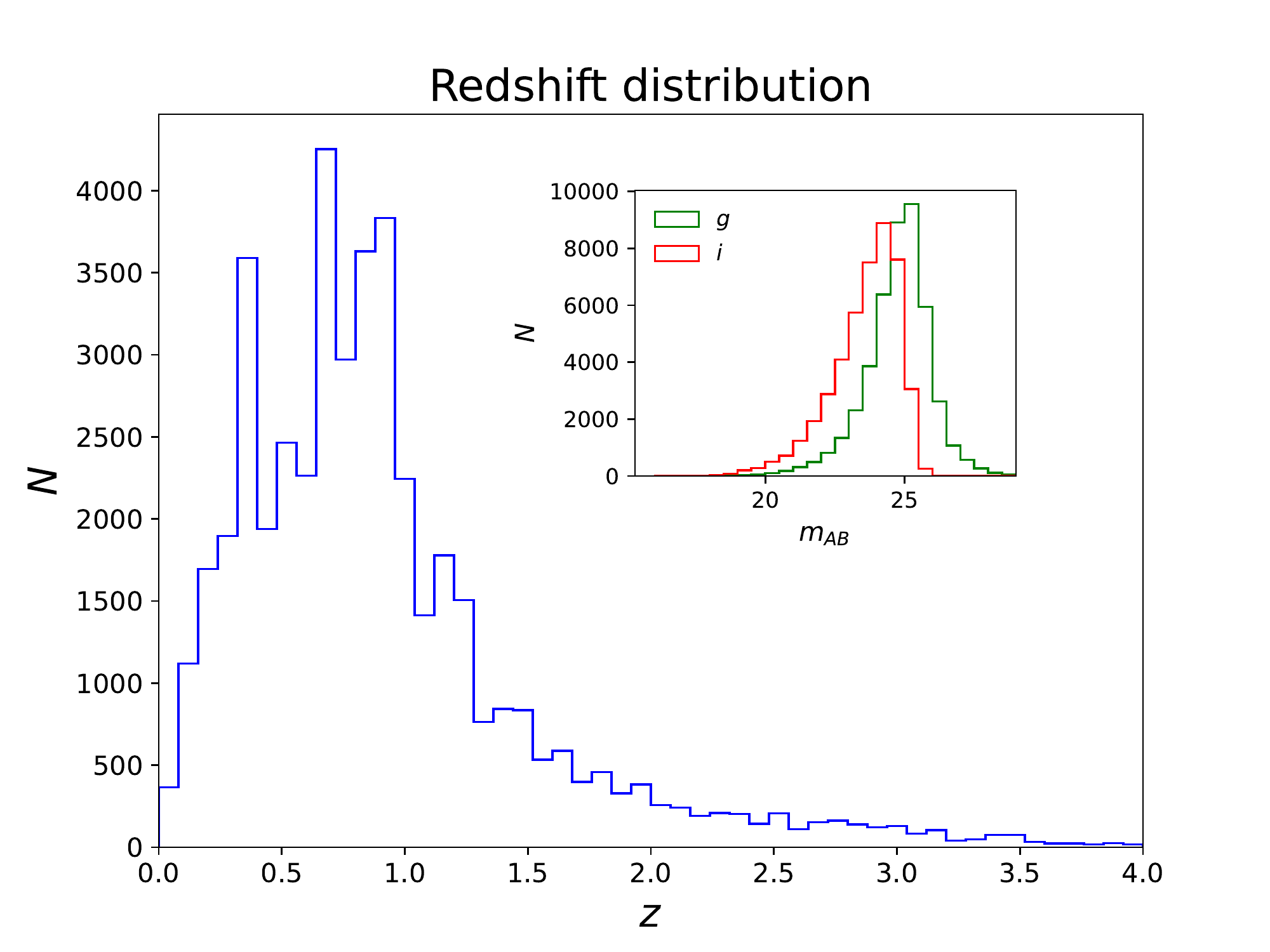}
  \caption{CSST galaxy redshift distribution derived from the COSMOS catalog. These galaxies are selected with SNR larger than
    10 in $g$ or $i$ bands. The distribution peaks around $z=0.6\sim0.7$, and can reach maximum at $z\sim4$. We also show the distribution of
    AB magnitudes in the $g$ and $i$ bands.}
  \label{fig:redshift_distribution}
\end{figure}

\begin{figure}
  \centering
  \includegraphics[width=\textwidth, angle=0]{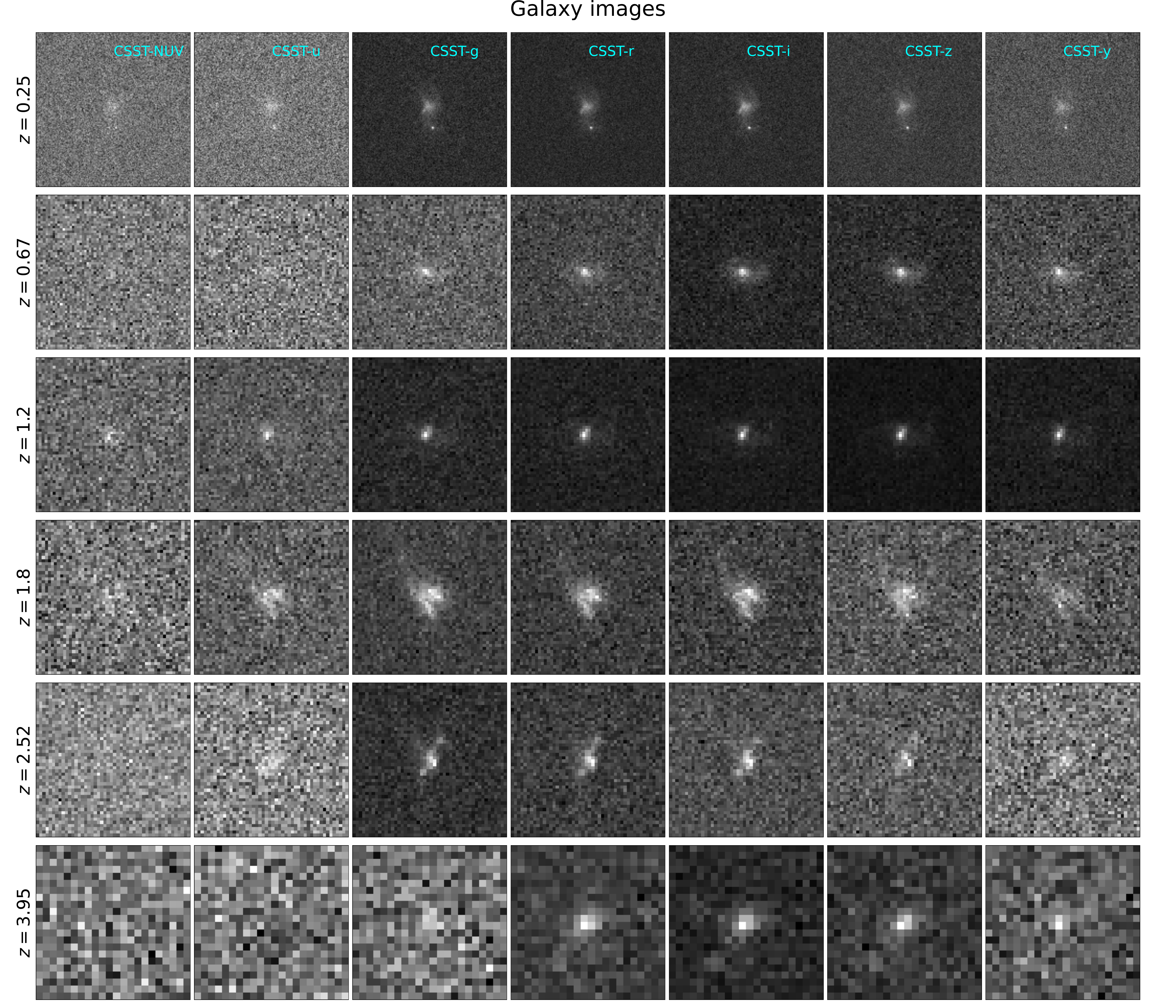}
  \caption{Examples of simulated galaxy sources in 7 CSST photometric bands at different redshifts.
    We notice that noises in $NUV, u$ and $y$ bands are more dominant since their transmissions
    are relatively low. Besides, some sources in high redshifts are almost overwhelmed by background noises in some band,
    and the neural network method can be applied to try to extract information in these images. }
  \label{fig:examples}
\end{figure}

\begin{figure}
  \centering
  \includegraphics[width=\textwidth, angle=0]{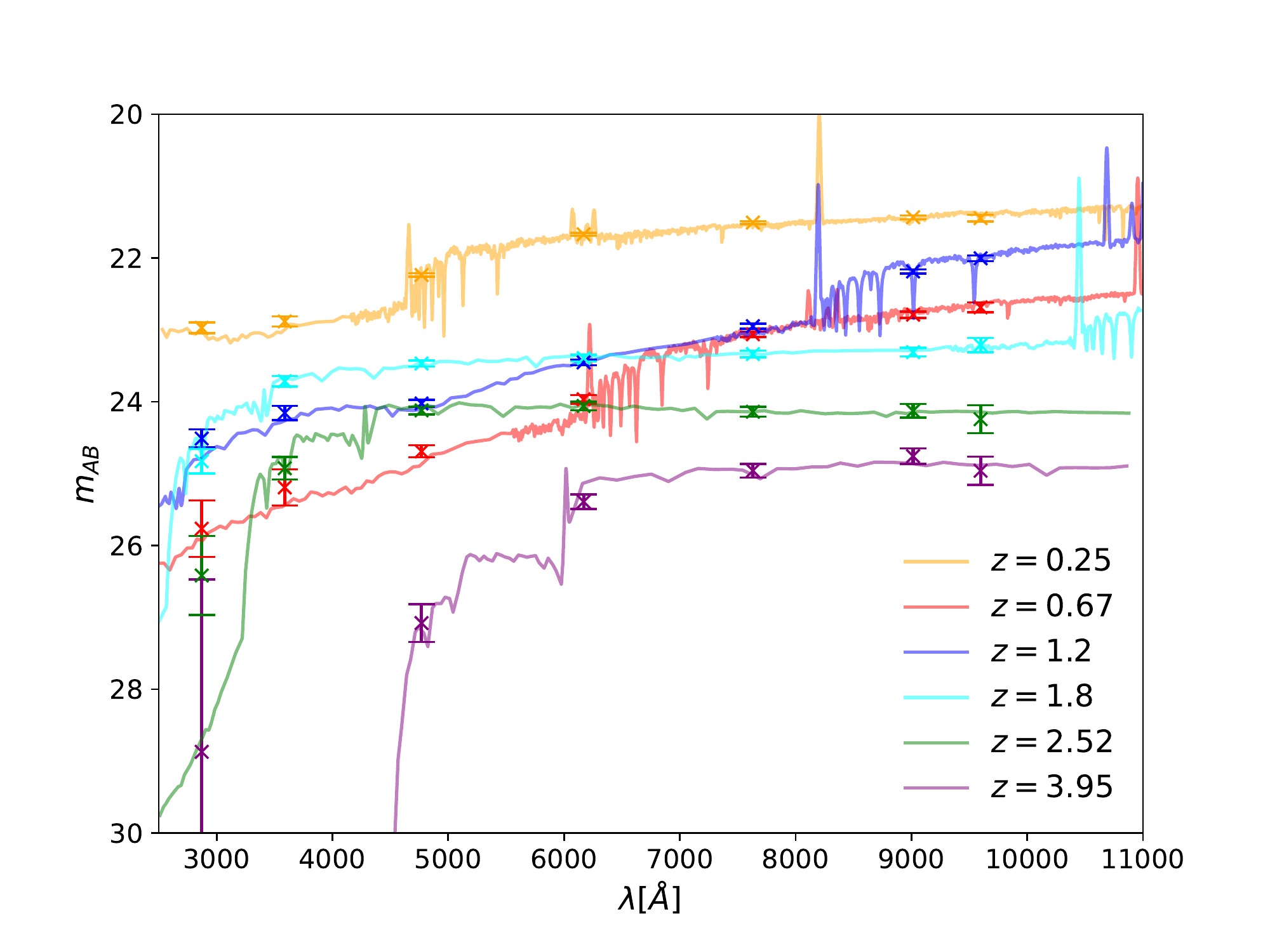}
  \caption{The corresponding fluxes of galaxy samples in Figure~\ref{fig:examples} measured by the aperture photometry method.
    The galaxy SEDs are also shown and rescaled to the levels of flux data for comparison.}
  \label{fig:SEDs}
\end{figure}

Firstly, we select an area of $0.85\times0.85$ deg$^2$ from the $HST$-ACS survey, where $\sim192,000$ galaxies can be identified.
Then we rescale the pixel size from $0.03$ arcsec of the $HST$ survey to $0.075$ of the CSST survey. The identified galaxies are extracted
as square stamp images with galaxies at the center of images. The image sizes are 15$\times$ galaxies' semi-major axis,
which can be obtained in the COSMOS weak lensing source catalog~\citep{Leauthaud2007}, so our galaxy images are in different sizes.
Other sources in the image are masked and replaced by background noise, and only the galaxy image in the center is reserved.

Then we can rescale galaxy images from the $HST$-ACS F814W survey to the CSST flux level by using galaxy SEDs to obtain the CSST 7-band images. Galaxy SEDs can be
produced by fitting fluxes and other photometric information given in COSMOS2015 catalog by $LePhare$ code~\citep{Laigle2016,Arnouts1999,Ilbert2006} and photo-$z$s from the catalog are fixed during the fitting procedure. The SED
templates applied are also from this catalog, and we extend these templates from $\sim900$ \AA\ to $\sim90$ \AA\ using the BC03 method
~\citep{Bruzual03} to include the fluxes of high-$z$ galaxies in all CSST photometric bands, where details can be found in ~\citet{Cao2018}. About $100,000$ high quality
galaxies with reliable photo-$z$ measurement are selected. And when fitting SEDs, we also consider dust extinction and emission
lines, such as Ly$\alpha$, H$\alpha$, H$\beta$, OII and OIII. After fitting the galaxy SEDs, we can calculate
the theoretical flux data by convolving with the CSST filter transmissions shown in Figure~\ref{fig:transmission}.
At the same time, fluxes of F814W images can be calculated
with aperture size of 2$\times$ of Kron radius~\citep{Kron1980}. Then, the CSST 7-band images can be produced by rescaling the fluxes.
The background noise also has to be adjusted to the same level of the CSST observation, and the details are given in~\citet{Zhou2022}.
After the noise is generated, we obtain the mock CSST galaxy images for the seven CSST photometric bands.

Our galaxy flux mock data are measured by aperture photometry. We first measure the Kron radius along major and minor-axis to obtain
an elliptical aperture of size $1\times R_{\rm Kron}$. Then the flux and error in each band can be calculate within this
aperture. Note that the measured fluxes in some bands could be negative due to relatively large background noise.
It has been proven that this effect does not affect training of our networks, and as shown later, we will rescale the fluxes
and try to reserve the information.

The redshift distribution of galaxy sources selected from the COSMOS catalog are shown in Figure~\ref{fig:redshift_distribution},
and the selection details are explained in the next section. Commonly, we need spectroscopic redshifts as accurate redshift values to train neural networks.
Here, we assume our selected galaxies can be seen as data with accurate redshifts or spec-$z$s, since we have fixed photo-$z$s from COSMOS2015 catalog
in our simulation procedure. Besides, since the purpose of this work is mainly about method validation, this assumption should be reasonable currently.
We can see that the distribution peaks around $z=0.6\sim0.7$, and can reach maximum at $z\sim4$, which is in consistency with previous studies~\citep{Cao2018,Gong2019,Zhou2021}.
Figure~\ref{fig:examples} shows some examples of CSST mock galaxy stamp images at different redshifts, and the corresponding SEDs of these galaxies are displayed in Figure~\ref{fig:SEDs}.
We notice that galaxies at low redshifts usually have higher SNRs with low backgrounds, contrarily, galaxies at high redshifts can be easily dominated by
background noise, especially in $NUV$, $u$ and $y$ bands with low transmissions. Thus neural network is necessary to be applied for extracting information
from these noisy images, and with Bayesian neural networks, uncertainties brought by background noise and the network itself can be well captured.

\section{Methods}
\label{sect:methods}
We use Bayesian MLP and Bayesian CNN to derive photo-$z$ from mock flux and image data respectively. And these two networks are
combined to test the improvement of accuracy when including both data. We first briefly introduce Bayesian neural networks,
and then the architectures and training process are discussed. All networks we construct are implemented by Keras\footnote{\url{https://keras.io}} with
TensorFlow\footnote{\url{https://tensorflow.org}} as backend and TensorFlow-Probability\footnote{\url{https://tensorflow.org/probability}}.

\subsection{BNN}
\label{sect:bnn}
Generally, neural network only produces point value estimate without errors, since weights are fixed after training and the output is simply the
values of parameters we are interested in. In order to correctly capture the parameter uncertainties, we have to understand where they come from.
Uncertainties brought by neural network compose of two parts.
One comes from instrinsic corruption of data, called aleatoric uncertainty, and this uncertainty can not be reduced in training
~\citep{Hora1996,Kiureghian2009}.
~\citet{Bishop1994} proposed Mixture Density Network (MDN) to capture aleatoric uncertainty using mixture of distributions to replace the point
estimate of networks. The output of MDN consists of the weights and parameters of each distribution,
say mean $\mu$ and standard deviation $\sigma$ for Gaussian distributions. After training, we can sample from the mixture of
distributions for testing data, and then calculate the uncertainty of predictions.

The other one is called epistemic uncertainty, which comes from insufficient training of network. Gathering more training data or taking average
of results from ensemble networks can reduce this kind of uncertainty. Bayesian network says the weights of network can be sampled from posterior
distributions learned by training data given proper priors of weights. Thus when testing, these weights vary in every run and the epistemic uncertainty
can be captured with enough runs.
Mathematically, we define prior of weights as $p(\omega)$ and the posterior of network weights learned from training data pair
${\mathbf{X, Y}}$ as $p(\omega|\mathbf{X, Y})$, so for test input $\mathbf{x}$, the distribution of output $\mathbf{y}$ can be
calculated as:
\begin{equation}
  \label{eq:inference}
  p(\mathbf{y|x, X, Y})=\int p(\mathbf{y|x}, \omega) p(\omega|\mathbf{X, Y})d\omega.
\end{equation}
The analytical calculation of this equation is difficult, since $p(\omega|\mathbf{X, Y})$ can not
be evaluated analytically. However, this distribution can be approximated by variational inference approach~\citep{Blundell2015}. In variational
inference, we define a variational distribution, $q(\omega)$, which has analytical form to replace $p(\omega|\mathbf{X, Y})$. The
parameters of this distribution are learned so that $q(\omega)$ is as close as possible to real posterior. The approximation can be
performed by minimizing their Kullback-Leibler (KL) divergence, which measures similarity between two distributions. Therefore,
Equation~\ref{eq:inference} can be rewriten as:
\begin{equation}
  \label{eq:inference with variational}
  p(\mathbf{y|x})\approx\int p(\mathbf{y|x}, \omega) q(\omega)d\omega.
\end{equation}
Minimizing KL divergence between $q(\omega)$ and $p(\omega|\mathbf{X, Y})$ is equivalent to maximizing the log-evidence lower bound
(log-ELBO)~\citep{Gal2015bayesian}, that is,
\begin{equation}
  \label{eq:log ELBO}
  \mathcal{L}_{VI} = \int q(\omega)\log{p(\mathbf{Y|X}, \omega)}d\omega - D_{\rm KL}(q(\omega)||p(\omega)),
\end{equation}
where the first term is the log-likelihood of output parameters of training data, and the second term can be
approximated as an $L_2$ regularization as shown in~\citet{Gal2015bayesian}. So this equation can be written as
\begin{equation}
  \label{eq:variational loss}
  \mathcal{L}_{VI}\approx \sum_{n=1}^{N}\mathcal{L}(\mathbf{y}_n, \mathbf{\bar{y}}_n(\mathbf{x}_n, \omega))
  - \lambda\sum_i|\omega_i|^2,
\end{equation}
where $n$ and $i$ denote number of training data and weights, respectively, and weights are sampled from $q(\omega)$.
$\mathcal{L}(\mathbf{y}_n, \mathbf{\bar{y}}_n(\mathbf{x}_n, \omega))$ is the likelihood of network prediction
$\mathbf{\bar{y}}_n(\mathbf{x}_n, \omega)$ for input $\mathbf{x}_n$ with labels $\mathbf{y}_n$, and $\lambda$ is the
regularization strength. Ignoring the regularization, minimizing KL divergence is to maximizing the log-likelihood.
After training, we can perform multiple runs of testing data through network to obtain output parameters mutiple times.
This procedure is identical to implementing Equation~\ref{eq:inference with variational} to sample from variational distribution $q(\omega)$ to construct
distributions of outputs, $p(\mathbf{y|x})$.

However, sampling from posterior distributions of weights only captures epistemic distributions. For regression tasks, if we
assume that the predictions of parameters of a single run also obey some distributions, then we can capture the aleatoric
uncertainties. Gaussian distributions are the most commonly used ones. Therefore, the log-likelihood
of Equation~\ref{eq:variational loss} can be written as a Gaussian log-likelihood:
\begin{equation}
  \label{eq:loss}
  \mathcal{L}(\mathbf{y}_n, \mathbf{\bar{y}}_n(\mathbf{x}_n, \omega))=\sum_{j}\frac{-1}{2\sigma_j^2}
  |y_{n, j}-\bar y_{n, j}(\mathbf{x}_n, \omega)|^2 - \frac{1}{2}\log\sigma_j^2,
\end{equation}
where $\sigma_j$ represents the aleatoric uncertainties of $j$th parameters inherited from corruption of input data. $\sigma_j$
can be predicted along with parameters. No labels of $\sigma_j$ are required, since they can be produced when balancing the two terms
in Equation~\ref{eq:loss}. Predictions with aleatoric uncertainties can be obtained by sampling from Gaussian distributions with
output parameters as means and standard deviations. Therefore, combining aleatoric and epistemic uncertainty is performing multiple
runs for testing data, and in each run, predictions are sampled from Gaussian distributions.

Bayesian neural networks are built upon special layers with trainable weights, where the forms of posterior and prior distributions must
be given. For simplicity, we select multivariate standard normal distribution as priors, thus the posteriors are normal distribution
with learnable means and deviations. In forward pass, the network samples weights from posteriors and estimates outputs from inputs.
However, backpropagation can not produce gradients of means and deviations from distributions. There is a trick to cope with this
problem called re-parameterization~\citep{Kingma2013}. This trick samples $\mathbf{\epsilon}$ from a parameter-free distribution and transforms $\mathbf{\epsilon}$
with a gradient-defined function $t(\mathbf{\mu, \sigma, \epsilon})$. Commonly, $\mathbf{\epsilon}$ is sampled from a standard normal distribution, i.e,
$\mathbf{\epsilon}\sim\mathcal{N}(0, \mathbf{I})$, and the function is defined as $t(\mathbf{\mu, \sigma, \epsilon})=\mathbf{\mu+\sigma\odot\epsilon}$,
which shifts the $\mathbf{\epsilon}$ by mean $\mathbf{\mu}$ and scales it by $\mathbf{\sigma}$, where $\odot$ is matrix element-wise
multiplication. Then the backpropagation algorithm can be accomplished. We use flipout layers introduced in~\citet{Wen2018},
which uses roughly twice as many floating point operations than re-parameterization layer, but can achieve significant lower
variance and speedup training process.

\subsection{Network Architecture}
\label{sect:network architecture}
\subsubsection{Bayesian MLP}
\label{sect:mlp}
Since our galaxy flux data consists of 7 discrete points obtained by 7 CSST photometric bands, we adopt MLP to predict photo-$z$ from flux data.
MLP is composed of input layers, hidden layers and output layers, and the layers are connected by trainable weights and biases. The internal
relationship between flux data and redshift can be learned when training. We apply DenseFlipout layer to build our Bayesian MLP.

More relevant information can give more accurate predictions, so our inputs of network are fluxes, colors and errors~\citep{Zhou2021}.
The fluxes and errors are typically in exponential form, and it is not suitable to directly input network, where the weights
will have very large fluctuations. To speed up training and reduce fluctuations of weights, data should be normalized or rescaled.
Therefore, we divide fluxes with corresponding fluxes of magnitude limits of 7 bands of CSST mentioned in Section~\ref{sect:intro}.
Note that our result is not sensitive to the divided values. And errors are
divided with their corresponding fluxes to obtain relative errors. Since colors are values of subtraction of magnitudes between
two bands, we construct color-like values as division of fluxes between two bands. There are still some large values inappropriate
for networks, so we need to rescale these values with a logarithmic function. As we mentioned, since fluxes and errors are measured
within apertures, some negative fluxes in bands severely affected by background noise may arise, especially in $NUV, u$ and $y$ bands.
To solve this problem, we use the following logarithmic function:
\begin{equation}
  f(x) = \left\{
  \begin{array}{lcc}
    \log(x)   &  & {x > 0,} \\
    -\log(-x) &  & {x < 0.} \\
  \end{array} \right.
  \label{eq:scale function}
\end{equation}
This function will rescale the fluxes, colors and errors obtained in the first rescaling, and reserve the negative information which
may be useful for photo-$z$ predictions. Hereafter, we simply call the rescaled fluxes, colors and errors as fluxes, colors and errors in the
context.

\begin{figure}
  \centering
  \includegraphics[width=\textwidth, angle=0]{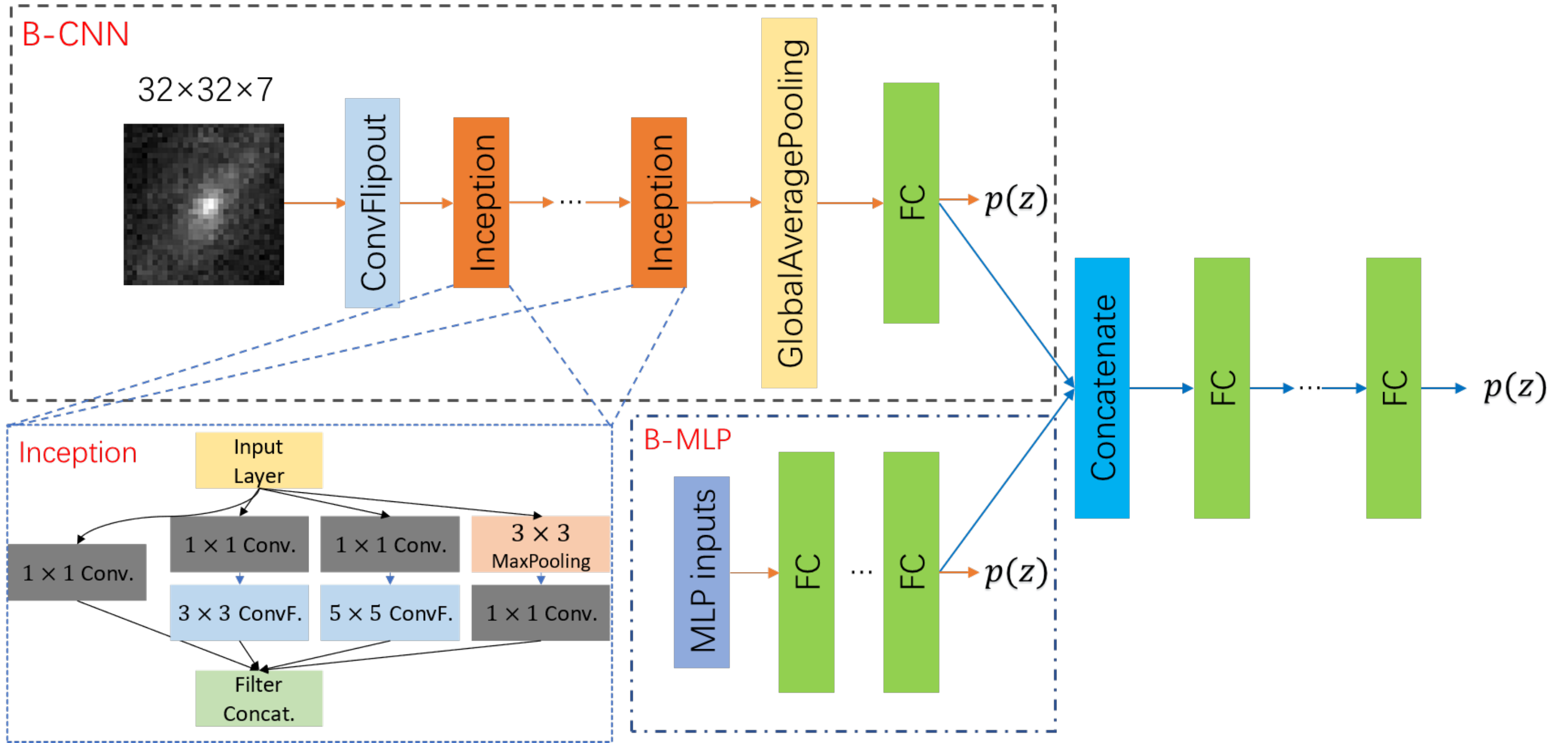}
  \caption{Architecture of our Bayesian MLP, CNN and Hybrid network. Bayesian MLP is shown in dash-dotted blue box. The inputs are fluxes, colors and errors,
    and 6 hidden or fully connected (FC) layers are stacked. The output is a Gaussian distribution of predicted photo-$z$.
    The dashed black box displays the structure of Bayesian CNN, and its input is $32\times32\times7$ images. The input is convolved by Convolution2DFlipout layer,
    and then downsampled, obtaining a feature map. Then the feature map is processed by 3 inception blocks, and the output is flattened to vector for
    connecting with following fully connected layer. Then the Gaussian distribution of photo-$z$ can be obtained. Inception block is illustrated in
    blue dashed box, where $3\times3$ and $5\times5$ kernels are used to extract features in different scales.
    Bayesian Hybrid network combines MLP and CNN by concatenating features extracted from them, and then
    several fully connected layers are structured to obtain the distribution of photo-$z$. Note that all layers with trainable weights are flipout layers,
    where the form of prior and posterior distribution must be provided, except for $1\times1$ convolution, since
    this layer simply acts as a scaling one to reduce channels of features and increase computation efficiency.
  }
  \label{fig:architecture}
\end{figure}

\renewcommand{\arraystretch}{1.5}
\begin{table}
  \caption{Details of Bayesian MLP architecture.}
  \label{tab:MLP_param}
  \begin{center}
    \begin{tabular}{lcc}
      \hline
      Layers                         & Output Status$^a$ & Number of params.$^b$ \\
      \hline
      \hline
      Input                          & 20                & 0                     \\
      \hline
      FC$^c$                         & 40                & 1640                  \\
      \hline
      ReLU                           & 40                & 0                     \\
      \hline
      FC$^c$                         & 40                & 3240                  \\
      \hline
      BatchNormalization             & 40                & 160$^d$               \\
      \hline
      ReLU                           & 40                & 0                     \\
      \hline
                                     & ...$^e$           &                       \\
      \hline
      Params                         & 2                 & 162                   \\
      \hline
      $\mathcal{N}(\mu, \sigma)$$^f$ & -                 & 0                     \\
      \hline
      \hline
    \end{tabular}
  \end{center}
  \vspace{-2mm}
  \textbf{Notes}.\\
  $^a$ Number of data points or neurons. \\
  $^b$ Total number of parameters: 18,802. \\
  $^c$ FC: fully connected layer. \\
  $^d$ Half of them are non-trainable parameters.\\
  $^e$ 4 repeats of FC + BatchNormalization + ReLU. \\
  $^f$ Output is a Gaussian distribution with mean $\mu$ and standard deviation $\sigma$ obtained in params. \\
\end{table}

\renewcommand{\arraystretch}{1.5}
\begin{table}
  \caption{Details of Bayesian CNN architecture.}
  \label{tab:CNN_param}
  \begin{center}
    \begin{tabular}{lcc}
      \hline
      Layers                         & Output Status$^a$ & Number of params.$^b$ \\
      \hline
      \hline
      Input                          & (32, 32, 7)       & 0                     \\
      \hline
      Convolution2DFlipout           & (16, 16, 32)      & 4064                  \\
      \hline
      LeakyReLU                      & (16, 16, 32)      & 0                     \\
      \hline
      Inception                      & (8, 8, 72)        & 17632                 \\
      \hline
      Inception                      & (4, 4, 72)        & 19552                 \\
      \hline
      Inception                      & (2, 2, 72)        & 19552                 \\
      \hline
      GlobalAveragePooling           & 72                & 0                     \\
      \hline
      FC$^d$                         & 40                & 5800                  \\
      \hline
      BatchNormalization             & 40                & 160$^c$               \\
      \hline
      ReLU                           & 40                & 0                     \\
      \hline
      Params                         & 2                 & 162                   \\
      \hline
      $\mathcal{N}(\mu, \sigma)$$^e$ & -                 & 0                     \\
      \hline
      \hline
    \end{tabular}
  \end{center}
  \vspace{-2mm}
  \textbf{Notes}.\\
  $^a$ Format: (dimension, dimension, channel) or number of neurons.\\
  $^b$ Total number of parameters: 66,922.\\
  $^c$ Half of them are non-trainable parameters.\\
  $^d$ FC: fully connected layer.\\
  $^e$ Output is a Gaussian distribution with mean $\mu$ and standard deviation $\sigma$ obtained in params. \\
\end{table}

Therefore, our Bayesian MLP has 20 inputs, i.e, 7 fluxes, 7 errors and 6 colors. We construct 6 hidden DenseFlipout layers with 40 units in each
layer~\citep{Zhou2022}, and find 6 hidden layers are proper to cope with this task. To reduce overfitting, after each layer except for the first,
BatchNormalization is applied~\citep{Ioffe2015}, and all layers are activated by Rectified Linear Unit
(ReLU) non-linear function~\citep{Nair2010}. Bayesian MLP outputs a Gaussian distribution constructed from redshift as mean and aleatoric
uncertainty as deviation. The details of architecture are shown in Table~\ref{tab:MLP_param} and in Figure~\ref{fig:architecture}.
Note that the parameters are approximately as twice large as non-Bayesian MLP shown in~\citet{Zhou2022}, since the weights are constructed by
Gaussian distributions with two parameters.

\subsubsection{Bayesian CNN}
\label{sect:cnn}
We use Bayesian CNN to predict photo-$z$ from CSST mock galaxy images. These images are from 7 CSST bands and can be considered as 2d-arrays with 7 channels,
from which 2d Bayesian CNN can extract information to predict photo-$z$. As we mentioned in Section~\ref{sect:mock data}, the images are sliced according
to semi-major axis of galaxies given in catalog, therefore, the final images are in different sizes. Since neural
network can only process data with same sizes, we need to crop images with size larger than a threshold area $S_{\rm threshold}$ and pad images
with size smaller than $S_{\rm threshold}$. Cropping is performed centrally since galaxies reside in center of images. And images are
padded with typical background noise derived in Section~\ref{sect:mock data} to better simulate the real observations.
The most proper value of $S_{\rm threshold}$ is proven to be $32$ pixels, and other sizes $16$ and $64$ pixels are also researched. We find smaller
one loses too much information since most galaxies occupy pixels larger than $16$, and the larger one introduces more background noise causing network
cannot concentrate on the central galaxies.

Inception blocks proposed by~\citet{Szegedy2014} can extract information in different
scales parallelly and effectively combine them. ~\citet{Pasquet2019} and ~\citet{Henghes2021} and our previous work build their networks
based on inception block to predict photometric
redshift from images and achieve quite accurate results. Therefore, we construct Bayesian inception blocks with flipout
layers. Our inception block is illustrated in Figure~\ref{fig:architecture} and use Convolution2DFlipout layers with $3\times3$ and $5\times5$ kernels to
extract features and learn the distributions of trainable weights. The $1\times1$ kernels can reduce channels of features
and increase computation efficiency and we do not use distributions to express the weights of these layers.

Our Bayesian CNN inputs $32\times32\times7$ images. The input images are firstly processed by a Convolution2DFlipout layer with $32$
kernels of $3\times3$ size and stride size $2$ to extract information and downsample images to $16$. Following first layer is $3$
inception blocks to learn more abstract features and we finally obtain feature images with a size $2$. In order to connect with fully connected
layer, we utilize global average pooling to vectorize the feature images to $72$ values~\citep{Lin2013} and employ one fully connected layer with
$40$ units. The fully connected layer is also built upon DenseFlipout with learnable distribution of weights.
Outputs of this network is a Gaussian distribution the same as Bayesian MLP mentioned above. Note that after each
Convolution2DFlipout layer, we apply BatchNormalization layer and ReLU activation function.
The details of architecture are shown in Figure~\ref{fig:architecture} and Table~\ref{tab:CNN_param}.
The parameters are about as twice large as non-Bayesian CNN shown in~\citet{Zhou2022}.

\subsubsection{Bayesian Hybrid}
\label{sect:hybrid}
The galaxy images in 7 bands abstractly contain fluxes, colors, errors and morphological information, which means these information
can be extracted from images. And our Bayesian CNN can directly learn photo-$z$s from images, probably extracts features related to these
information. In contrast, our MLP inputs the obvious flux information and fitting the relation between redshifts and flux data. Hence,
if we combine MLP and CNN to construct a hybrid network and this network can input both flux data and images, it may result in more
accurate photo-$z$ predictions. We construct hybrid network by concatenating
Bayesian MLP and CNN mentioned above in both last fully connected layer, obtaining a vector of size $80$, and then structure $6$ fully
connected layers with $80$ units built upon DenseFlipout to learn distribution of weights. And after each layer, BatchNormalization and
ReLU activation function are applied. The output of this hybrid network is the same as Bayesian MLP and CNN mentioned above. The
schematic diagram is illustrated in Figure~\ref{fig:architecture}.

\subsection{Training}
\label{sect:training}
Here, we follow \citet{Zhou2022} and select
about $40,000$ high-quality sources with SNR in $g$ or $i$ band larger than $10$ from generated dataset mentioned in Section~\ref{sect:mock data}.
These sources all possess both flux data and images. As we notice, we consider their reliable photo-$z$s as accurate spectroscopic redshifts that
can be used in our training process. And
in the real CSST survey, samples with spec-$z$ obtained by future deep spectroscopic surveys will be used to retrain our networks.
The above samples are divided into training and testing sets.
We spare $10,000$ samples for testing and the rest, about $30,000$ are used for training, constructing a ratio of training to
testing to be approximately $3:\ 1$. And we split 10\% for validation from training data.
We also try $1:\ 1$ and $1:\ 3$ ratio to study the influence of training size on accuracy of photo-$z$.

Our Bayesian MLP uses the negative of Gaussian log-likehood (negative of Equation~\ref{eq:loss}) as our loss function,
which considers both aleatoric and epistemic uncertainties.
The Adam optimizer is adopted to optimize weights of the network. This optimizer can adjust learning rate of every weight automatically given an initial
learning rate, which is set to be $10^{-4}$ for this network. We create an accuracy metric based on the definition of outlier percentage
to be $|z_{\rm true}-z_{\rm pred}|/(1 + z_{\rm true}) < 0.15$. Note that this metric cannot represent the final result, since the $z_{\rm pred}$ is a
random draw from learned distribution of photo-$z$.
This accuracy metric is monitored in training as well as negative log-likelihood loss. The maximum number of epoch and batch size in each epoch
are set to be $2000$ and $2048$, respectively.  In order to reduce the statistical noise and create more data, we augment training data by random
realizations based on
flux errors with Gaussian distribution~\citep{Zhou2022}. Here $50$ realizations are created and more realizations cannot significantly improve the results.
In training, we notice that the validation accuracy and loss follow the training ones well, and no overfitting occurs.

Our Bayesian CNN uses the same loss function and optimizer. The initial learning rate is also set to be $10^{-4}$. The maximum number of
epoch is also $2000$. Batch size in every
epoch is set to be $1024$. We save the model when validation loss and accuracy are converged.
We augment training data by including their rotated and flipped counterparts,
resulting a $8\times$ data size. This augmentation can probably make the network more accustomed to background noise and better concentrate on
central galaxies.

Bayesian hybrid network uses the same setting of loss and optimizer. The
maximum number of epoch and batch size is $2000$ and $512$ and the converged model with steady validation loss and accuracy is saved as our final model.
The inputs of MLP part are augmented by $50$ random realizations based on errors,
and the images are randomly rotated or flipped to correspond
one specific realization of flux data. Thus the network can reduce the statistical noise brought by flux data, and can be accustomed to background noise
at the same time.

Since Bayesian MLP and CNN can predict photo-$z$ accurately, the features learned by the two networks are optimized to fullfill this task. To explore if
further improvement exists for photo-$z$, we try to investigate that if the features learned by both CNN and MLP are better than features directly
learned by hybrid network. We create a hybrid transfer network, inspired by the techniques from transfer learning (TL), which utilizes existing
knowledge from one problem to solve related problem~\citep{Pan2009}. The MLP and CNN parts of this network is transferred from trained ones and their
weights are frozen.
Thus the features combined by this network are the ones already learned by MLP and CNN, respectively, instead of directly learned by the hybrid network.
Note that the structure of this network is the same as the hybrid one, except for employing a new training strategy from transfer learning. In training,
we find freezing the layers before the last fully connected layers of MLP and CNN to include more flexibility result better.

\subsection{Calibration}
\label{sect:calibration}

The uncertainties predicted by neural network are probably miscalibrated~\citep{Guo2017, Ovadia2019}. We can examine if our network is
well calibrated through reliability diagram, which shows the coverage probability of samples with true values residing in specific
confidence interval. If the true values of $x\%$ samples lies in $x\%$ confidence interval, then this network model is well
calibrated~\citep{Levasseur2017, Hortua2020} and the reliability diagram should be a straight diagonal line.

The calibration can be achieved by tuning hyper-parameters of networks when training, such as the kernel size for convolution layer,
regularization parameters and so on. However, fine-tuning hyper-parameters is a challenging and time-consuming task. Calibration after
training is also an option and the relevant methods are described in literature (see references in~\citet{Hortua2020}). We use
Beta calibration introduced in~\citet{Kull2017}. Firstly, we construct the reliability diagram for testing data and use following function
to fit this line:
\begin{equation}
  \beta(x; a, b, c) = \frac{1}{1 + 1/(e^c\frac{x^a}{(1 - x)^b})}
  \label{eq:fitting function}
\end{equation}
where $a, b$ and $c$ are the fitting parameters. We scale covariance matrix $\Sigma$ by a factor $s$ to obtain $s\Sigma$, and choose the $s$ parameter for minimizing the difference between the fitting line and diagonal line. Therefore, $s\Sigma$ is a well calibrated covariance matrix. For our photo-$z$ work, we just need to scale the uncertainty of every sample.

\section{Results and Discussions}
\label{sect:results}
\begin{figure}
  \centering
  \includegraphics[width=\textwidth]{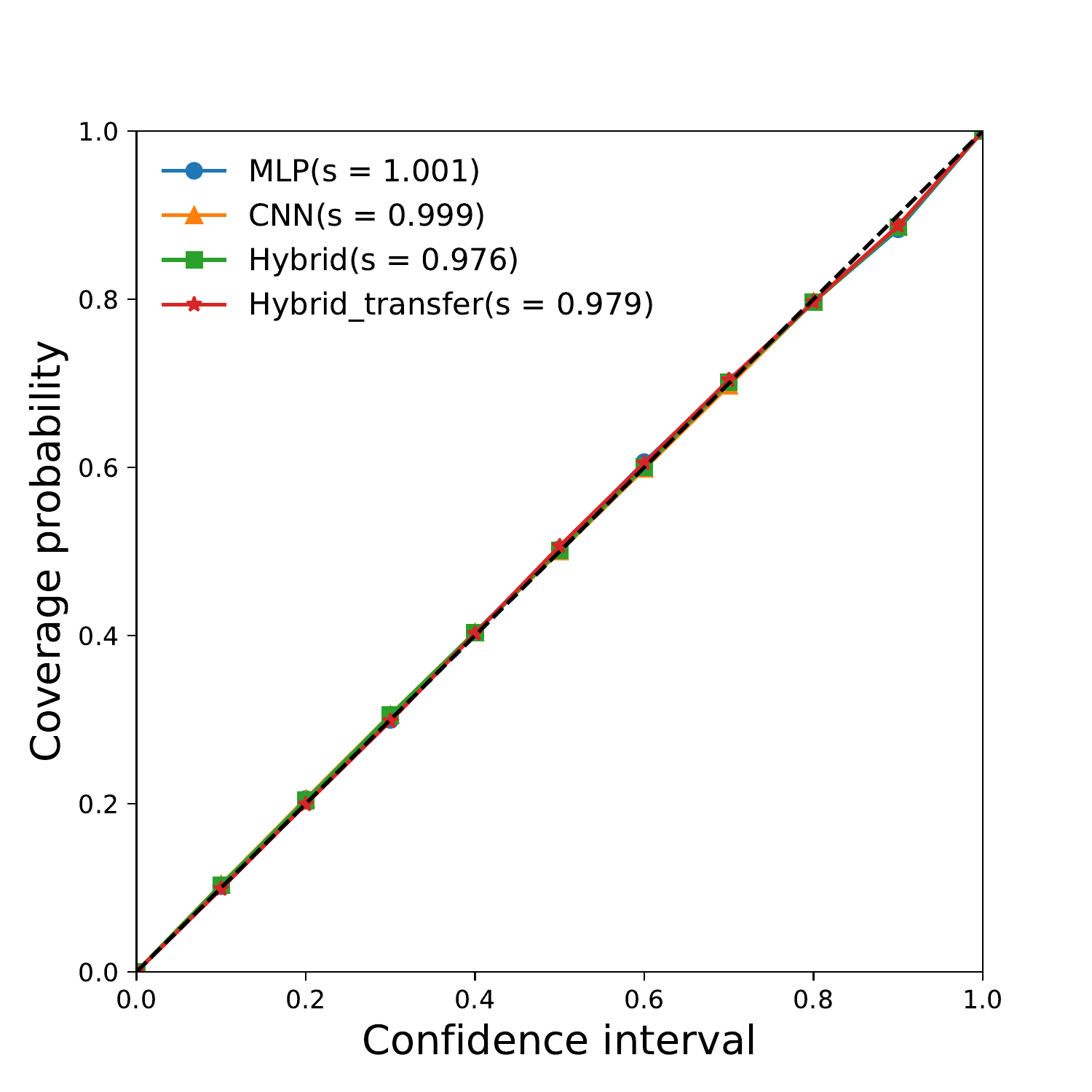}
  \caption{The reliability diagram for four networks. The uncertainties predicted by four networks are reliable when calibrated after Beta
    calibration method. }
  \label{fig:calibration}
\end{figure}

The percentage of catastrophic outliers is widely used in photo-$z$ research. Here, we define our catastrophic outliers to be
$|\delta z|/(1+z_{\rm true})<0.15$, where $\delta z = z_{\rm pred}-z_{\rm true}$. The normalized median absolute deviation is another mainly adopted
quantity, which can be calculated as
\begin{equation}
  \label{eq:sigma}
  \sigma_{\rm NMAD} = 1.48\times{\rm median}\left(\left|\frac{\delta z - {\rm median}(\delta z)}{1 + z_{\rm true}}\right|\right).
\end{equation}
This deviation shows the scattering of predictions considering the evolution of redshift, and provides a proper estimation of accuracy.
The average of $|\delta z|/(1+z_{\rm true})$ as MAE is also calculated for comparison with literature. Besides, to measure the performance of confidence,
we calculate the average 1$\sigma$ photo-$z$ uncertainties $\overline{E}$ in the predictions. And we define a similar metric called coverage introduced
in~\citet{Jones2022} to examine our reliability of uncertainties:
\begin{equation}
  \label{eq:coverage}
  C = \sum_i^{N_{gal}}\frac{|\overline{z}_{{\rm pred}, i} - z_{{\rm true}, i}| < \sigma_i}{N_{gal}}
\end{equation}
where $N_{gal}$ is the number of galaxy samples in specific redshift bin, and $\sigma$ is the 68\% confidence interval of prediction for every sample.
After feeding testing data to our four networks, we calibrate these models with the method mentioned in Section~\ref{sect:calibration} and plot a reliability
diagram in Figure~\ref{fig:calibration}. Notice that they are well calibrated and the scaling parameters are close to 1, which means that they are almost self-calibrated when training.


\begin{figure}
  \centerline{
    \includegraphics[scale=0.26]{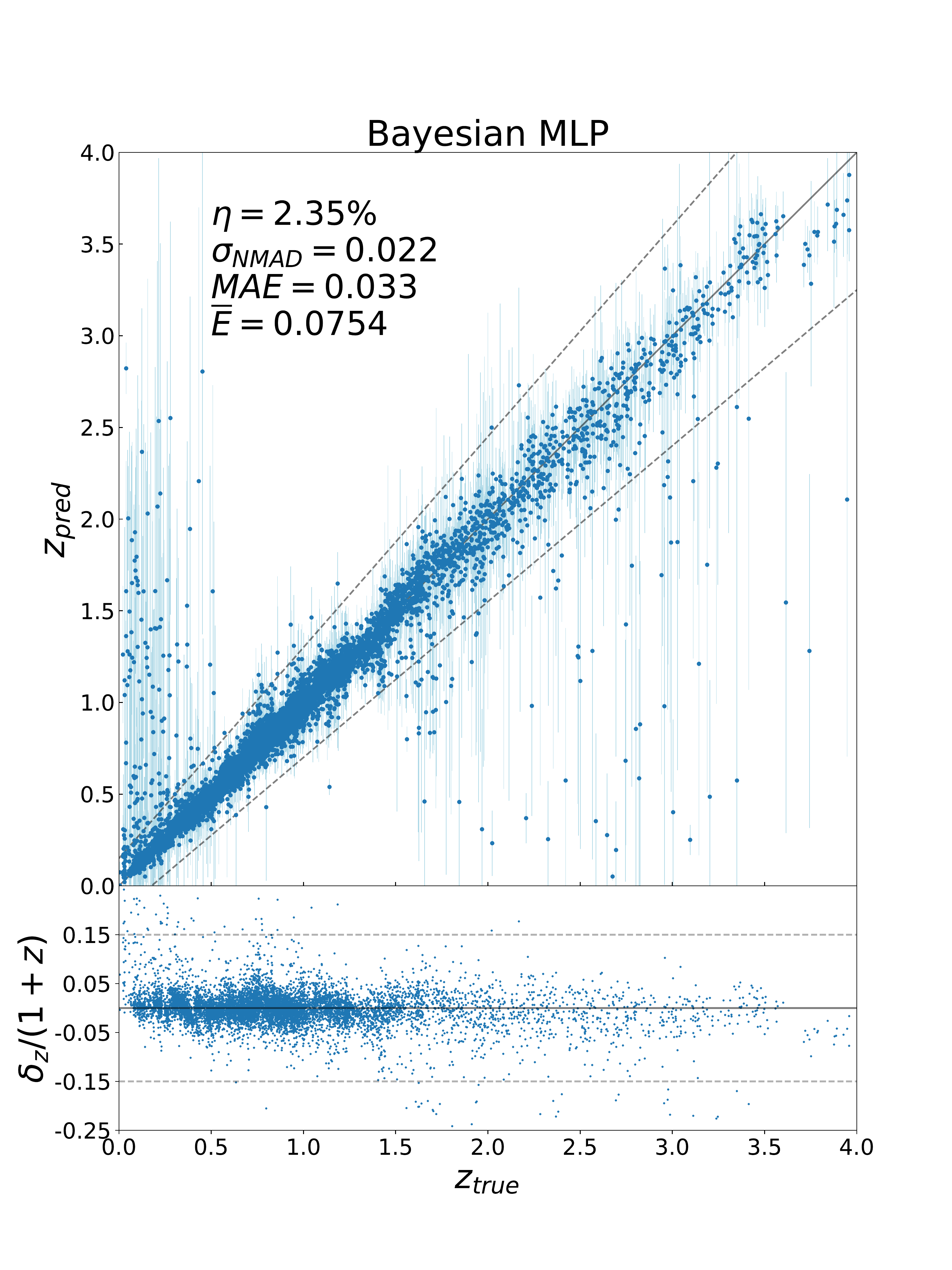}
    \includegraphics[scale=0.26]{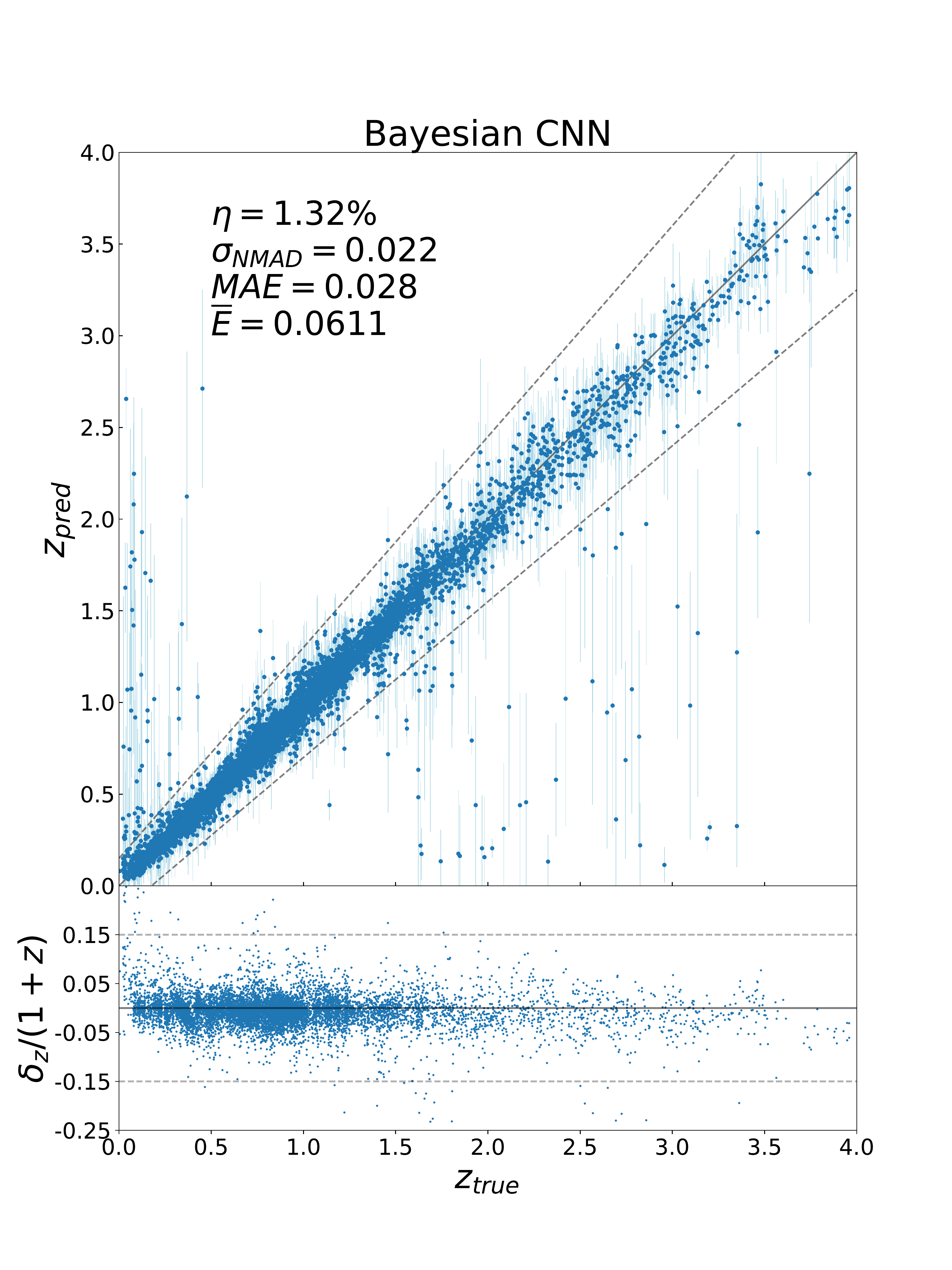}}
  \centerline{
    \includegraphics[scale=0.26]{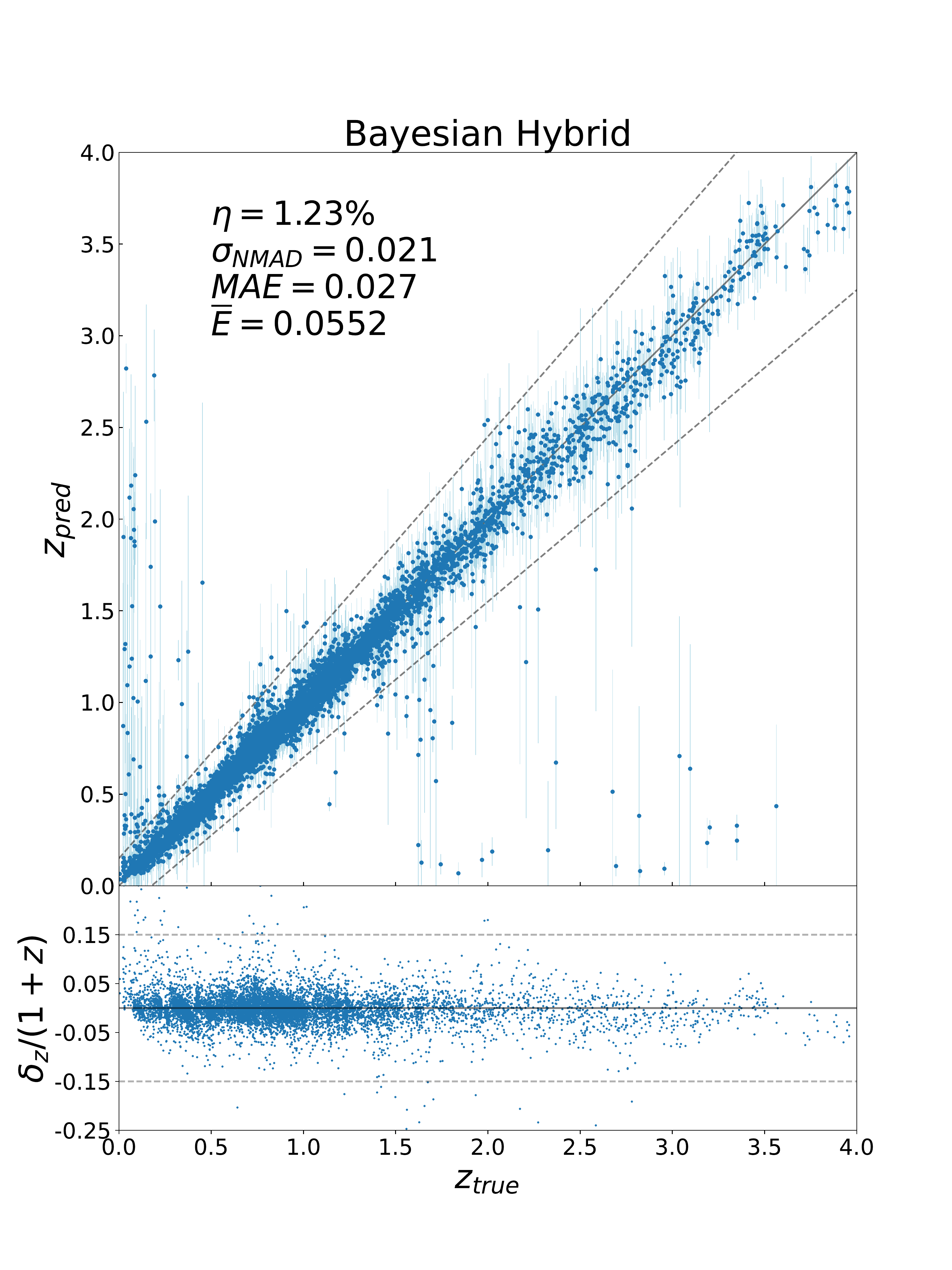}
    \includegraphics[scale=0.26]{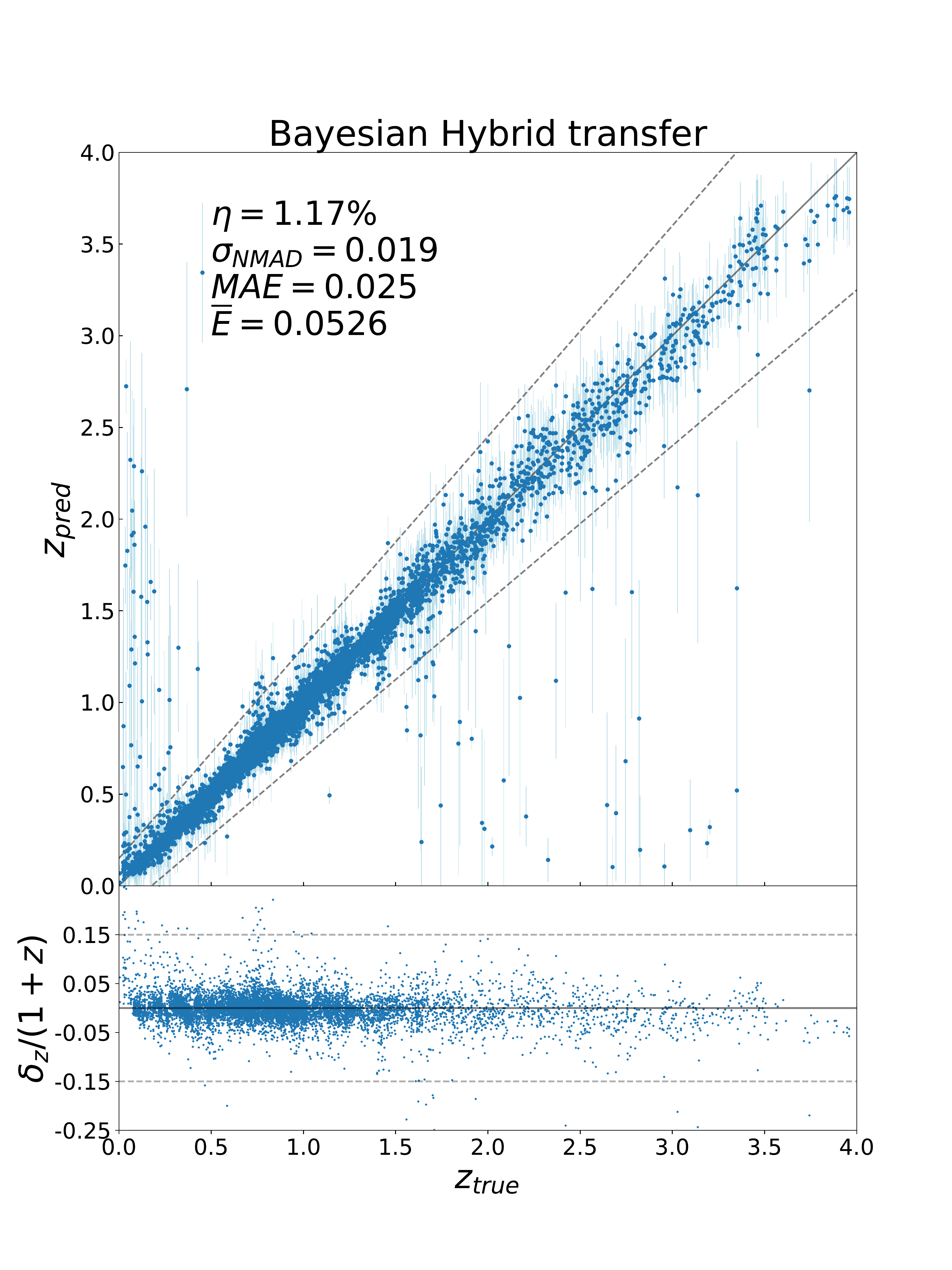}}
  \caption{Photo-$z$ result of Bayesian MLP, CNN, hybrid and hybrid transfer networks. The $\eta, \sigma_{\rm NMAD}$
    and $\overline{E}$ represents the outlier fraction, normalized median absolute deviation, and average 1$\sigma$ uncertainties or errors, respectively.
    The error bars are derived from the Gaussian distributions output by the networks. Hybrid and hybrid transfer networks can achieve outlier fraction
    smaller than 1.5\% and $\sigma_{\rm NMAD}\simeq0.02$.}
  \label{fig:photoz result}
\end{figure}

\begin{figure}
  \centering
  \includegraphics[width=\textwidth, angle=0]{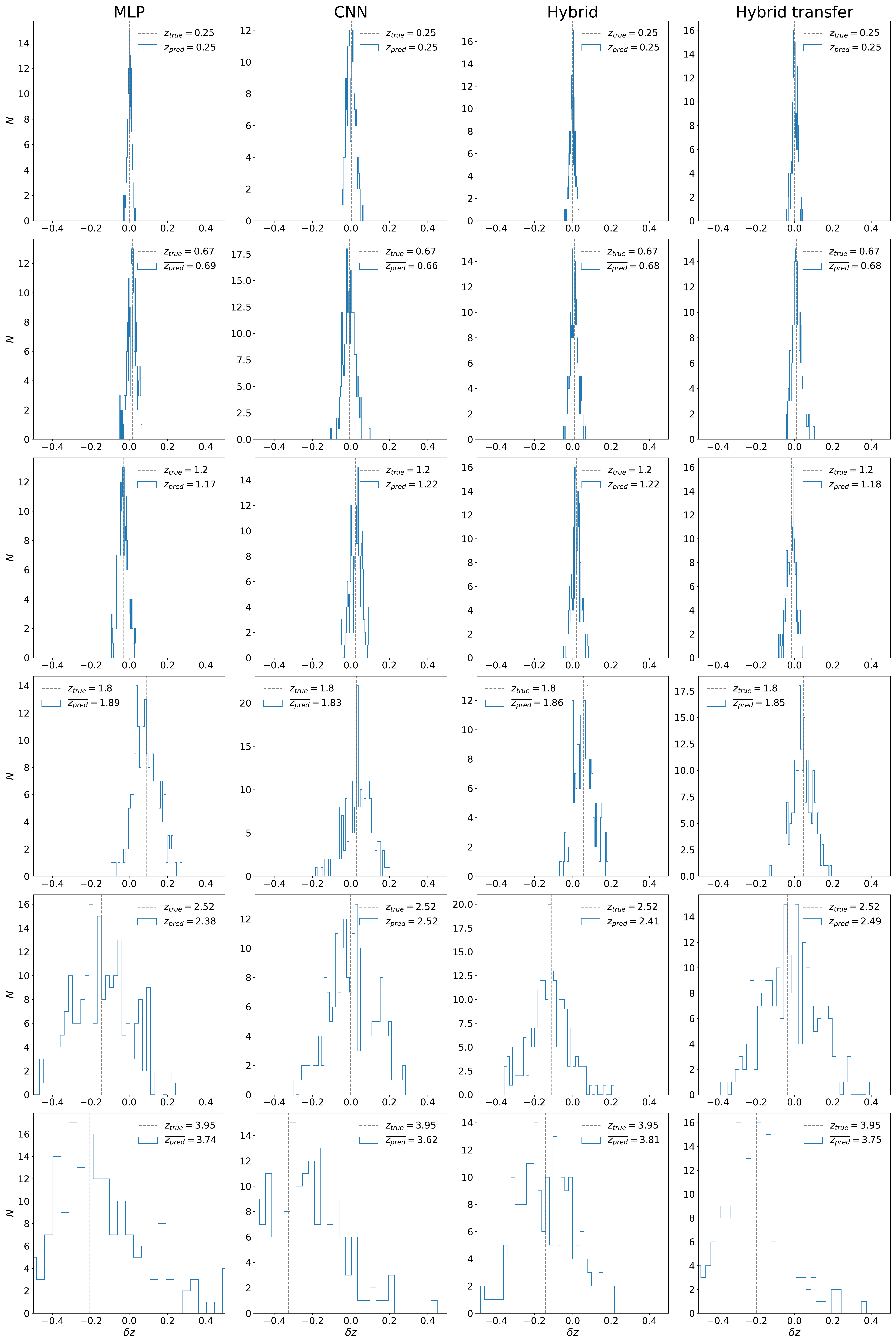}
  \caption{PDFs for the deviations of predicted and true redshifts, provided by the Bayesian MLP, CNN, hybrid and hybrid transfer network for the galaxy samples shown in Figure~\ref{fig:examples}. Dashed lines show the deviations of average predictions to the true redshifts.}
  \label{fig:PDFs}
\end{figure}

\begin{figure}
  \centering
  \includegraphics[width=\textwidth, angle=0]{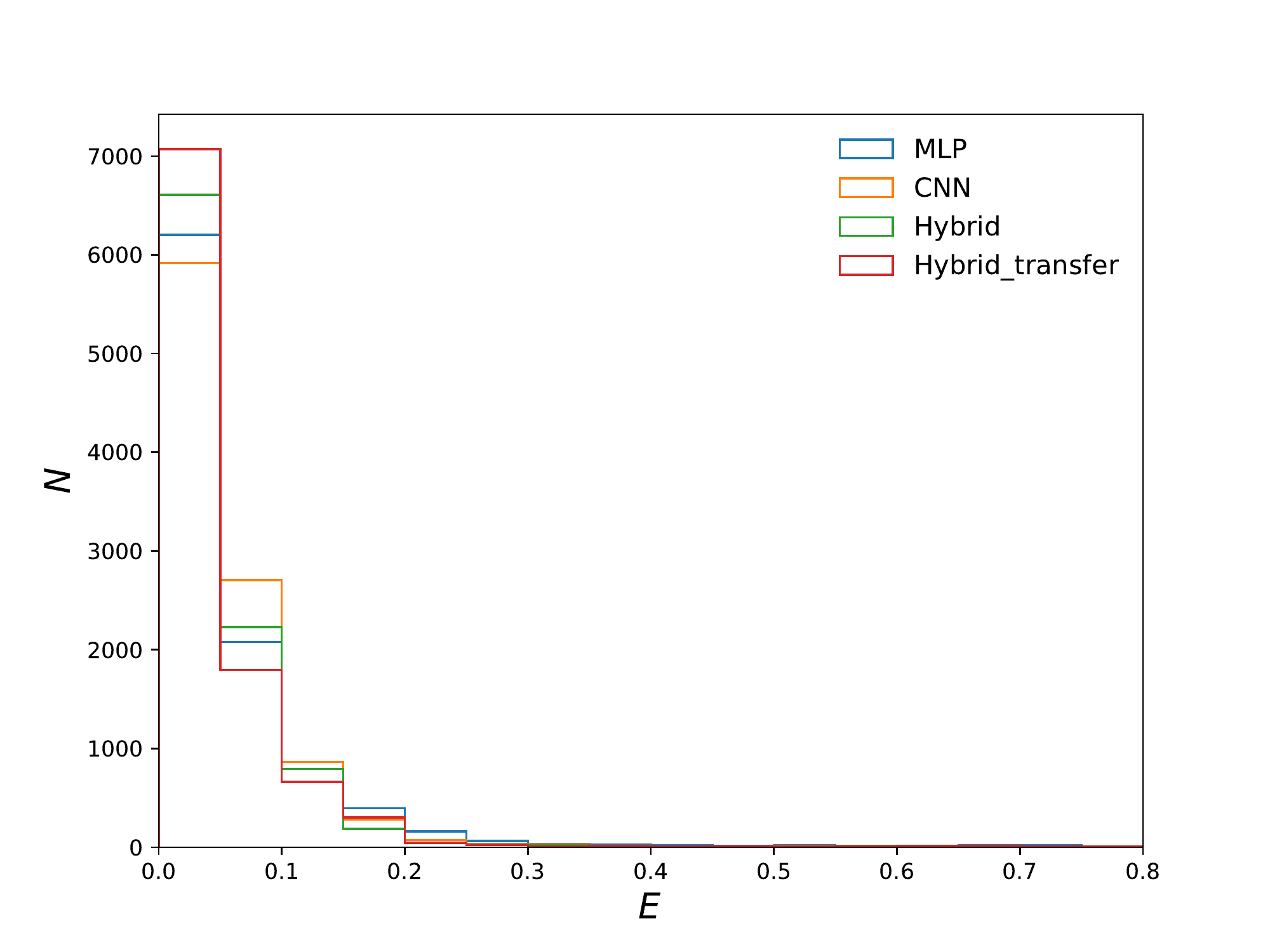}
  \caption{The distributions of photo-$z$ uncertainties for the four networks. Most uncertainties are lower than 0.2, but the ones of MLP
    can be higher, reaching a maximum at about 1.5.}
  \label{fig:uncertainty distributions}
\end{figure}

\begin{figure}
  \centering
  \includegraphics[width=\textwidth, angle=0]{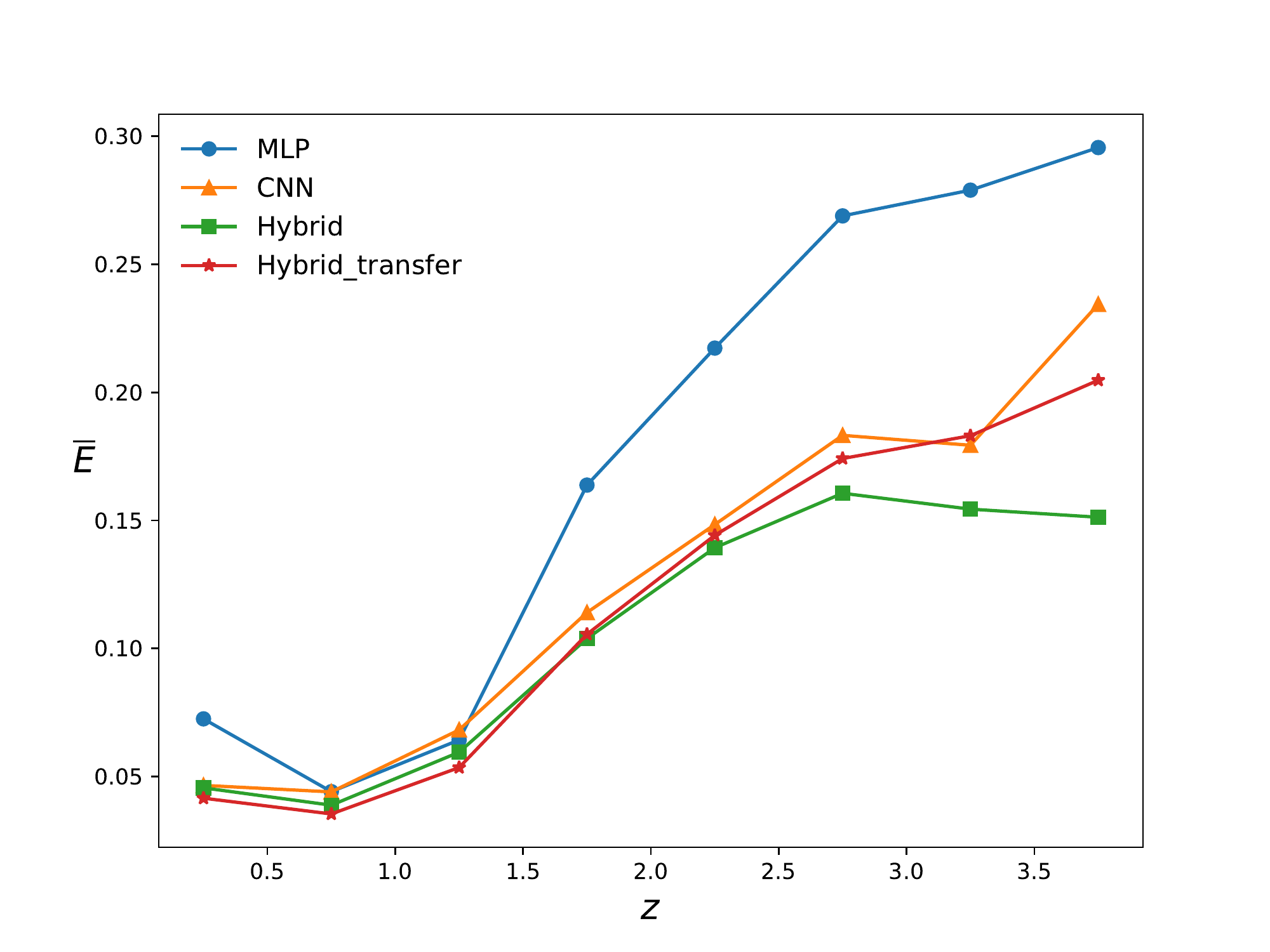}
  \caption{The average photo-$z$ uncertainties in different redshift bins. The four networks perform similarly well in redshift range $0.5\sim1.5$, assuring accuracy and confidence for most of galaxies.}
  \label{fig:average uncertainty}
\end{figure}

\begin{figure}
  \centering
  \includegraphics[width=\textwidth]{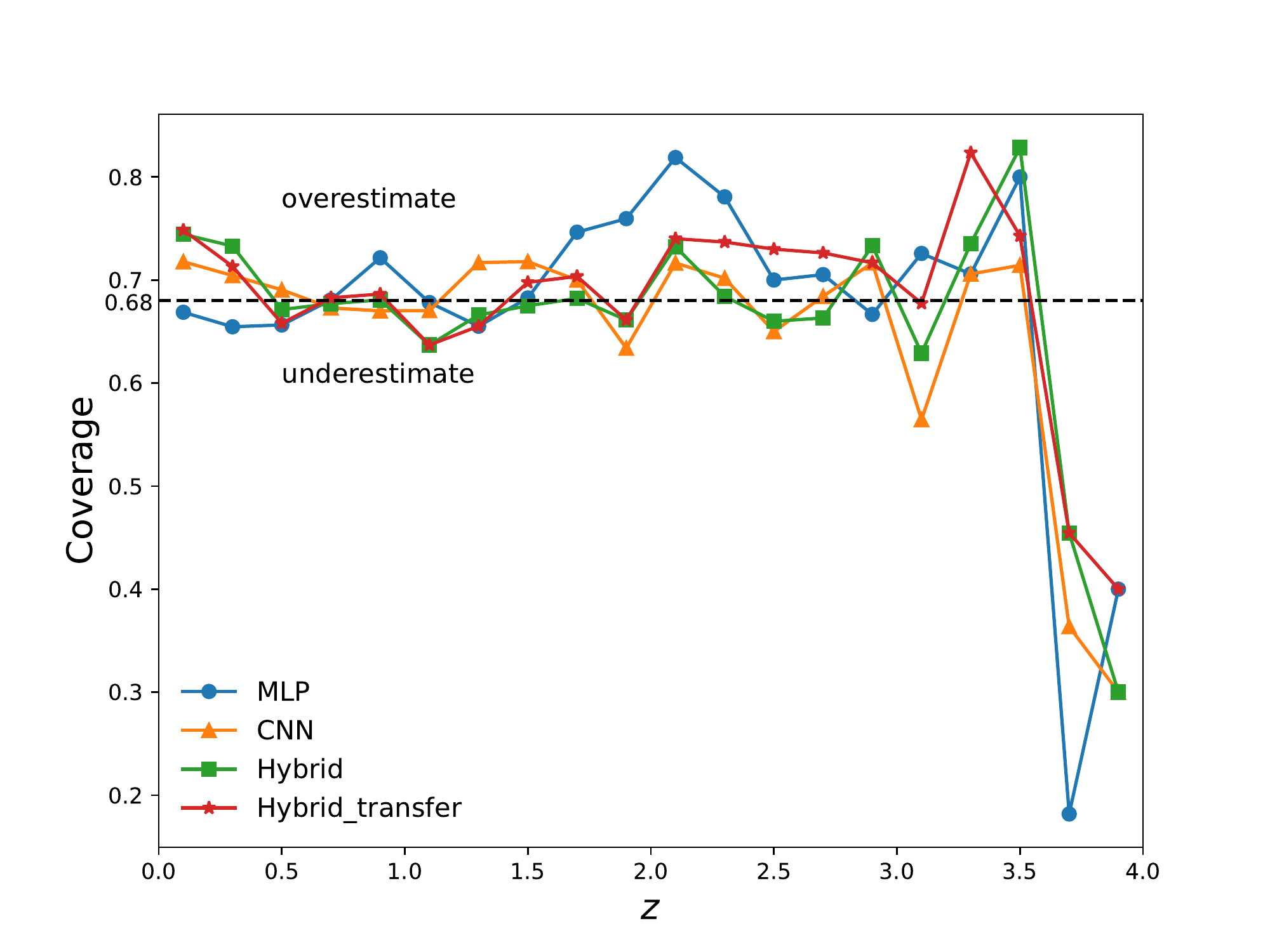}
  \caption{The coverage of photo-$z$ predictions. We notice that our curves fluctuate around 0.68 in low redshifts, and the fluctuations become larger at higher redshifts. Uncertainties of samples with $3.5 < z < 4.0$ are highly underestimated, probably resulting from statistical variance with few samples.}
  \label{fig:coverage}
\end{figure}

We show the photo-$z$ result for Bayesian MLP using the flux mock data in the upper-left
panel of Figure~\ref{fig:photoz result}, finding that the outlier fraction is 2.35\% and $\sigma_{\rm NMAD}$ is 0.022 for testing data.
The outlier fraction is higher than the estimation ($\sim1.4\%$) using the normal MLP
in~\citet{Zhou2022}, since much larger number of trainable parameters are included in the Bayesian MLP which are probably more difficult to optimize
and suppress the prediction accuracy. Maybe the price of obtaining the uncertainties of predictions is worse performance in point estimate. However,
we obtain similar $\sigma_{\rm NMAD}$ as ~\citet{Zhou2022}, meaning the dispersion of predictions is at the same level. The error bars are large for
some samples with redshift lower than $0.5$ and larger than $1.5$, meaning that network cannot predict these redshift well and assign them with very high
uncertainty. However in $0.5 < z < 1.5$, the outlier fraction and uncertainties are much lower, assuring high accuracy and confidence for most of galaxies
observed by CSST. This feature is probably due to the number of training data used at different redshifts, that most galaxies in the training sample are
within $0.5 < z < 1.5$ (see Fig.~\ref{fig:redshift_distribution}).

The upper-right panel of Figure~\ref{fig:photoz result} shows the result from Bayesian CNN using galaxy images. We find that the Bayesian CNN can
achieve outlier fraction 1.32\% and $\sigma_{\rm NMAD}=0.022$.
The outlier fraction is better than the MLP result. Since
images should abstractly include both morphological and flux information, in principle, the CNN could potentially extract all of these information, and provide
comparable or even better predictions than the MLP using the flux information only.

Lower panels of Figure~\ref{fig:photoz result} shows the result of Bayesian hybrid and hybrid transfer result, respectively.
The outlier fraction and $\sigma_{\rm NMAD}$ are 1.23\% and 0.021, and 1.17\% and
0.019 for the two networks. We can see that combining flux data and galaxy images can further decrease the outlier fraction. And the one employing transfer
learning provides slightly better result, implying features from trained MLP and CNN are probably more proper than
features directly learned by the hybrid network. The performance by hybrid and hybrid transfer networks are obviously better than the MLP and CNN, which
means that properly including both morphological and flux information can improve photo-$z$ predictions.

Figure~\ref{fig:PDFs} displays the PDFs for the deviation of predicted and true redshifts, provided by the Bayesian MLP, CNN, hybrid and
hybrid transfer networks, respectively, for the galaxy samples given in Figure~\ref{fig:examples}. Dashed lines show the deviations of average
photo-$z$ predictions to the true redshifts. We notice that at high redshift, the results are highly deviated from 0 and the PDFs are much wider,
resulting less confident predictions.

Figure~\ref{fig:uncertainty distributions} shows the distributions of predicted photo-$z$ uncertainties for the four networks. We notice that most uncertainties
stay below 0.2, but the ones of MLP can be
higher, reaching a maximum at about 1.5. We also show the average photo-$z$ uncertainties $\overline{E}$ in different redshift bins in
Figure~\ref{fig:average uncertainty}.
Here the bin size we use
is 0.5. The uncertainties for the four networks in redshift range from 0.5 to 1.5 are similarly small, assuring the accuracy for most galaxies.
The MLP are relatively higher in the whole redshift range, explaining the messy plot in upper-left panel in Figure~\ref{fig:photoz result}.
The uncertainties of CNN, hybrid and hybrid transfer networks are suppressed compared to the MLP case, and hybrid transfer
achieves the lowest uncertainties in low redshifts where most galaxies reside, but hybrid succeed in higher redshifts.
We calculate the average photo-$z$ uncertainties $\overline{E}$ for the whole range, and we have $\overline{E}$ = 0.754, 0.0611, 0.0552 and 0.0526 for Bayesian MLP, CNN,
hybrid and hybrid transfer networks, respectively. We note that the hybrid and hybrid transfer networks result in similar average uncertainties, and hybrid transfer performs slightly better.

We also plot the ``coverage" metric originally defined in~\citet{Jones2022}, which examines reliability of uncertainties with redshifts. In
Figure~\ref{fig:coverage}. We notice our curves fluctuate around 0.68 in low redshifts and the fluctuations become larger at higher redshifts.
Uncertainties of samples with $3.5 < z < 4.0$ are highly underestimated, probably resulting from statistical variance with few samples.

\renewcommand{\arraystretch}{1.5}
\begin{table}[]
  \centering
  \caption{Result comparison for our networks trained with different training data size.}
  \label{tab:result_comp}
  \begin{tabular}{|c|c|c|c|c|c|}
    \hline
    Train size(train-test ratio)                                             & statistics          & MLP    & CNN    & Hybrid & Hybrid transfer \\ \hline
    \multirow{3}{*}{\begin{tabular}[c]{@{}c@{}}30,000\\ (3: 1)\end{tabular}} & $\sigma_{\rm NMAD}$ & 0.022  & 0.022  & 0.021  & 0.019           \\ \cline{2-6}
                                                                             & $\eta$              & 2.35\% & 1.32\% & 1.23\% & 1.17\%          \\ \cline{2-6}
                                                                             & $\overline{E}$      & 0.0754 & 0.0611 & 0.0552 & 0.0526          \\ \hline
    \multirow{3}{*}{\begin{tabular}[c]{@{}c@{}}20,000\\ (1: 1)\end{tabular}} & $\sigma_{\rm NMAD}$ & 0.022  & 0.022  & 0.021  & 0.019           \\ \cline{2-6}
                                                                             & $\eta$              & 2.48\% & 1.64\% & 1.41\% & 1.28\%          \\ \cline{2-6}
                                                                             & $\overline{E}$      & 0.0758 & 0.0594 & 0.0578 & 0.0532          \\ \hline
    \multirow{3}{*}{\begin{tabular}[c]{@{}c@{}}10,000\\ (1: 3)\end{tabular}} & $\sigma_{\rm NMAD}$ & 0.023  & 0.024  & 0.023  & 0.021           \\ \cline{2-6}
                                                                             & $\eta$              & 2.43\% & 1.81\% & 1.67\% & 1.44\%          \\ \cline{2-6}
                                                                             & $\overline{E}$      & 0.0794 & 0.0656 & 0.0610 & 0.0551          \\ \hline
  \end{tabular}
\end{table}

\begin{figure}
  \centering
  \includegraphics[width=\textwidth]{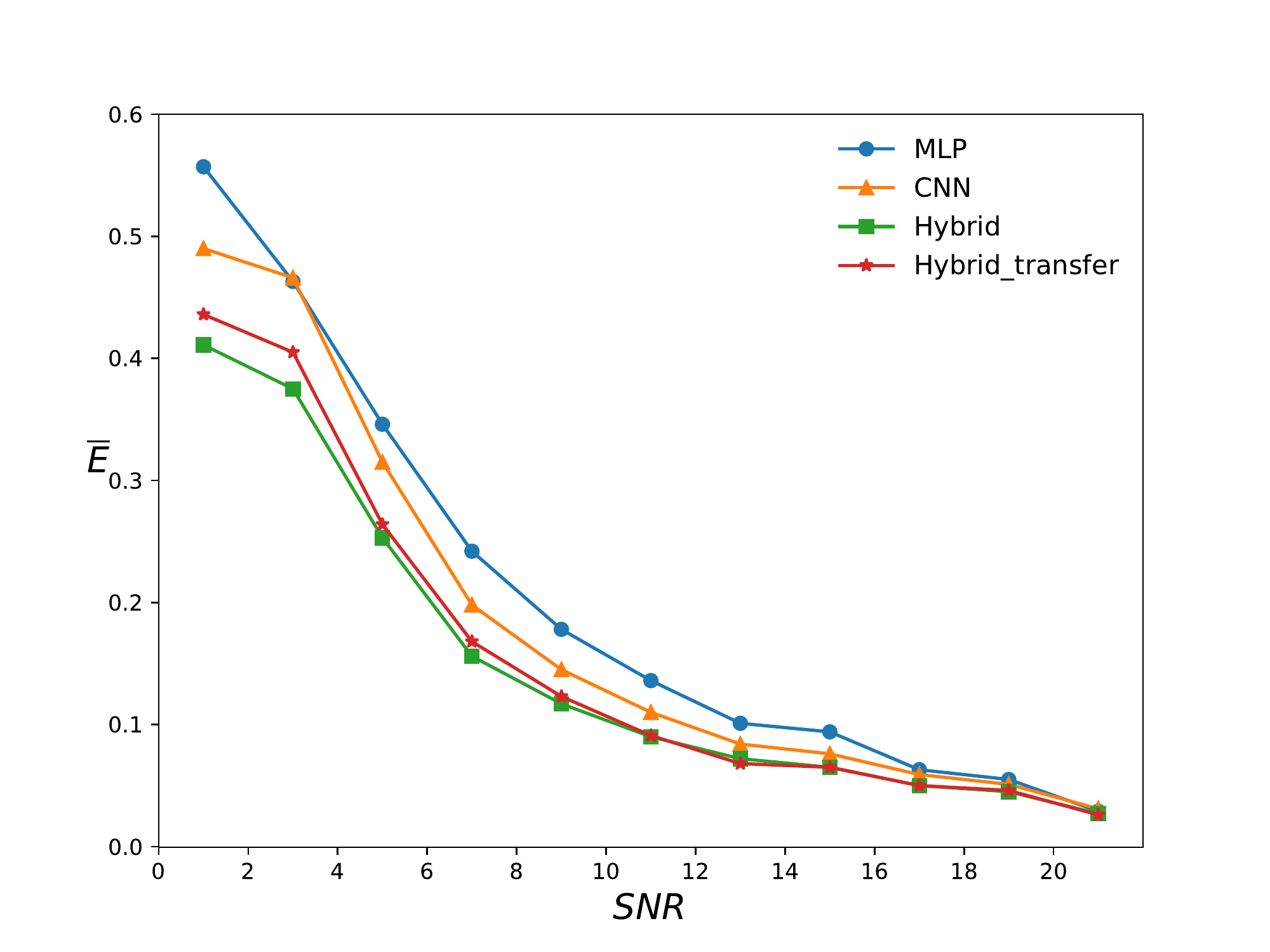}
  \caption{The relation of average uncertainties and SNR. We notice that as expected, the average uncertainties
    decrease when the SNRs become larger, and hybrid and hybrid transfer networks perform better than MLP and CNN case.}
  \label{fig:uncertainty_to_snr}
\end{figure}

Note that the results above are analyzed based on a training sample with $\sim$30,000 galaxies (a training and testing ratio of about $3:\ 1$ in our case).
We also test if the performance of the four networks will
be severely affected when feeding smaller set of training data, since we probably do not have large number of high quality photometric samples with
spectroscopic redshifts in real observations. Here, we split the data so that the training data
are about 20,000 (train-test ratio of $3:\ 1$) and 10,000 (train-test ratio of $1:\ 3$) to retrain these networks and calculate the results,
which is shown in~Table \ref{tab:result_comp}. And the calibration scale parameters $s$ are 0.998, 0.963, 1.140, 1.048 and 1.105, 0.935, 1.370, 0.882 for MLP,
CNN, hybrid and hybrid transfer networks trained with 20,000 and 10,000 data respectively. Decreasing training data does not provide severely worse results. We notice that MLP
even improves its outlier percentage from 20,000 to 10,000, probably because 10,000 training data is enough for training of MLP. The hybrid transfer network is
robust to decreasing of training data, providing similarly confident predictions.

In order to investigate the relationship between the SNR and uncertainties of photo-$z$, we select data with SNR in $g$ or $i$ band larger than 1, sparing 20,000 for testing, and retrain all networks. We show the average uncertainty as a function of the SNR in Figure~\ref{fig:uncertainty_to_snr}. We notice that the uncertainties
decrease with SNR growing as expected, and the hybrid and hybrid transfer networks perform better than the MLP and CNN case. Hybrid transfer results are worse in lower SNR, probably because
of the influence of transferred MLP and CNN parts, and they reach similar level when the SNR is larger than 10. The last points converge because they are calculated for samples with SNR$>$20.

\section{Conclusion}
\label{sect:conclusion}
In this work, we use Bayesian neural networks to explore the photo-$z$ accuracy and uncertainty for the CSST photometric survey.
The CSST data is simulated based on the COSMOS catalogs. Here we use four networks, including Bayesian MLP, CNN, hybrid and hybrid transfer networks.
The Bayesian framework is built upon variational inference technique, so that the weights are posterior distributions learned from given prior
and training data. The distributions of weights account for epistemic uncertainty, which comes from insufficient training and lack of data. On the other hand,
the aleatoric uncertainty coming from intrinsic corruption of data also needs to consider.

Bayesian MLP inputs flux data, including flux, color and error. These inputs are all rescaled to proper value range to speed up the training process.
Bayesian CNN processes galaxy images from the seven CSST bands. Our CNN is built upon inception blocks, which can extract information in different scales and
is beneficial for predicting photo-$z$.
Bayesian hybrid and hybrid transfer network are combinations of MLP and CNN through their learned features. The hybrid transfer network shares the same
architecture with the hybrid
networks, but applies a different training strategy borrowed from transfer learning. This hybrid transfer network freezes the weights of MLP and CNN parts
transferred from trained ones, and only the latter layers are optimized.

We find that all of these networks can derive accurate photo-$z$ results and use calibration method to obtain reliable uncertainties of predictions.
CNN can provide lower outlier fraction and more confident predictions than the MLP,
indicating CNN is capable to extract more information from the images besides the flux data. The hybrid and hybrid transfer networks result in similar
performance with the hybrid transfer slightly outperforming in outlier fraction and average photo-$z$ uncertainty. This result shows that feeding network with
both flux data and images can improve the photo-$z$ predictions. We also explore the effect of decreasing training samples, finding that smaller samples do
not severely corrupt the predictions in our case. The relationship between SNR and uncertainties of photo-$z$ is studied, and as expected, the average uncertainties decrease with SNR increasing.

We also should note that the BNNs actually need more optimization and are more time-consuming compared to traditional neural networks, since the BNNs usually have more tunable weights and the training process is more complex. However, as our work indicates, the BNNs are quite suitable and useful in photo-$z$ estimation that can obtain reliable uncertainties and PDFs with similar photo-$z$ accuracy as traditional neural networks. This means that the BNN should be a powerful tool and has large potentials to be applied in astronomical and cosmological studies.

\normalem
\begin{acknowledgements}
  We thank Nan Li for helpful discussions.  X.C.Z. and Y.G. acknowledge the support of MOST-2018YFE0120800, 2020SKA0110402, NSFC-11822305, NSFC-11773031, NSFC-11633004, and CAS Interdisciplinary
  Innovation Team. X.L.C. acknowledges the support of the National Natural Science Foundation of China through grant No. 11473044, 11973047, and the
  Chinese Academy of Science grants QYZDJ-SSW-SLH017, XDB 23040100, XDA15020200. L.P.F. acknowledges the support from NSFC grants 11933002, and the
  Dawn Program 19SG41 \& the Innovation Program 2019-01-07-00-02-E00032 of SMEC. This work is also supported by the science research grants from the
  China Manned Space Project with NO.CMS-CSST-2021-B01 and CMS- CSST-2021-A01. This work was funded by the National Natural Science Foundation of China (NSFC)
  under No.11080922. The numpy, matplotlib, scipy, tensorflow and keras python packages are used in this work.
\end{acknowledgements}

\bibliographystyle{raa}
\bibliography{bibtex}

\end{document}